\documentclass[12pt]{article}
\usepackage{bbold,amsmath,latexsym,multirow,epsfig}
\def\beq{\begin{equation}}
\def\eeq{\end{equation}}
\def\bea{\begin{eqnarray}}
\def\eea{\end{eqnarray}}
\def\nn{\nonumber}
\def\mc{\mathcal}
\def\tr{\textrm{Tr}}
\def\I{\mathbb{1}}
\def\t{\widetilde}
\def\o{\omega}
\def\ob{{\overline\omega}}
\def\ot{\widetilde\omega}
\def\obt{\widetilde{\overline\omega}}
\def\p{{\bf{p}}}
\def\q{{\bf{q}}}

\def\wt{\widetilde{w}}
\def\mt{\widetilde{m}}
\def\W{{\mathcal{W}}}

\def\ih{{\overline{I}}}
\def\jh{{\overline{J}}}

\def\O{{\mathcal{O}}}

\newcommand\dW[1]{{\frac{\delta\W}{\delta{#1}}}}
\newcommand\ddW[2]{{\frac{\delta^2\W}{\delta{#1}\delta{#2}}}}

\title{
\vspace{0.0 cm}
{\huge
An Equivalent Gauge and the \\ \vspace*{5pt} 
 Equivalence Theorem
}
\vspace{0.4cm}
\author{{\Large\text{Andrea Wulzer}\footnote{andrea.wulzer@pd.infn.it}} \\ 
{\small\emph{Dipartimento di Fisica e Astronomia, Universit\`a di Padova and}}\\ \vspace{-21pt} \\
{\small\emph{INFN, Sezione di Padova, via Marzolo 8, I-35131 Padova, Italy
}}}
}

\date{}

\begin{document}
\baselineskip=16pt

\maketitle \thispagestyle{empty}
\begin{center}
{\large Abstract} \\

\vspace{1cm}
\parbox[c]{12cm}{
I describe a novel covariant formulation of massive gauge theories in which the longitudinal polarization vectors do not grow with the energy. Therefore in the present formalism, differently from the ordinary one, the energy and coupling power-counting is completely transparent at the level of individual Feynman diagrams, with obvious advantages both at the conceptual and  practical level.

\medskip
\noindent
Since power-counting is transparent, the high-energy limit of the amplitudes involving longitudinal particles is immediately taken, and the Equivalence Theorem is easily demonstrated at all orders in perturbation theory. Since the formalism makes the Equivalence Theorem self-evident, and because it is based on a suitable choice of the gauge, we can call it an ``Equivalent Gauge''.
}

\end{center}
\newpage

\section{Introduction}

Massive gauge theories are so extraordinarily important for the physics of Fundamental Interactions that studying them requires no specific motivation, provided one can make some little progress on such a well-understood subject. However we do have a specific reason to be particularly interested in the high energy regime of these theories, which is now, and for the first time, under direct experimental investigation at the LHC. With the first run of the LHC machine, and even more so with the planned upgrade to $14$~TeV, particle physics is entering the age of high energy EW processes. In the forthcoming years it will become more and more common to deal with such processes, both from the experiment and from the theory side. 

In the high energy regime, where highly boosted EW bosons are involved, the standard formulation of massive gauge theories suffers of a well-known technical limitation, which I aim to overcome in the present paper. This is the fact that the polarization vectors associated with longitudinally polarized vector bosons, of energy $E$ and mass $m$, display and anomalous high energy behavior, namely they grow like $E/m$. This growth is problematic because it often does not correspond to a physical effect, in most cases the extra powers of $E$ from the polarization vectors cancel out in the final result, and this frequently happens through a complicated conspiracy among different diagrams. The longitudinal $W$ scattering process, $W_LW_L\,\to\,W_LW_L$ is a famous example of this situation. By naive power-counting, taking into account the energy behavior of vertices and propagators, one would predict at high energy a quartically divergent scattering amplitude, ${\mc{A}}\sim g_W^2(E/m)^4$, while the actual result is much different. In the absence of a Higgs boson,  ${\mc{A}}\sim (E/v)^2$, where $v=2\,m/g_W$ is the EWSB scale, while in the SM ${\mc{A}}\sim\lambda$, where $\lambda$ is the quadrilinear Higgs coupling.\footnote{A second contribution, of order $g_W^2$, is also present in the SM amplitude, the quadrilinear coupling contribution dominates only for  heavy Higgs.} Naive power-counting badly fails for this process. Not only it predicts the wrong energy scaling, but also the wrong dependence on the couplings. In spite of originating from gauge vertices, with coupling proportional to $g_W$, the $W_L$ scattering is not mediated by the gauge force, but by completely different interactions. Indeed the amplitude would remain different from zero also in the limit $g_W\to0$. This happens because $m=g_Wv/2$, and therefore the $E/m$ factors from the polarization vectors carry negative powers of $g_W$ which cancel positive powers from the Feynman vertices and change the coupling dependence of the final result.\footnote{The top quark decay, which is mediated by the Yukawa and not by the gauge force, is probably the most famous textbook  example of this phenomenon.}

In summary, the $E/m$ behavior of the polarization vectors invalidates power-counting, and this is a limitation in all problems where a coupling or energy expansion needs to be set up. At the purely theoretical level, the problem shows up when one tries to demonstrate general theorems for high energy EW processes, related for instance to high energy factorization like the Effective $W$ Approximation (EWA) \cite{Dawson:1984gx}. In that context, the lack of a reliable power-counting makes the theorem virtually impossible to prove in the standard covariant gauges, to the point that some authors \cite{Kleiss:1986xp} were led to question its validity. The EWA is on the contrary rather straightforward to demonstrate in the axial gauge \cite{Kunszt:1987tk,Borel:2012by}, where there is no anomalous growth of the polarization vectors and power-counting is manifest. However the axial gauge is not covariant, for the EWA and for similar applications it could instead be useful to formulate a covariant gauge where the polarization vectors are well-behaved, in this way one would combine the advantage of explicit covariance with the one of manifest power-counting. Identifying one gauge with these properties is the purpose of the present paper.

For what concerns phenomenology, the lack of a reliable power-counting in the ordinary covariant gauges is also a problem. It makes difficult to understand the physical origin of a given effect and to estimate its size before preforming an explicit calculation. This is a problem already in the SM, but even more so in the context of BSM theories, where plenty of new coupling are typically introduced. A quick estimate of their effects, or deciding whether or not they are relevant in the high-energy limit, is mandatory. Power-counting would also help with explicit calculations, either at the tree-level or including radiative corrections. By power-counting one would be able to select the relevant Feynman diagrams at a given order in a coupling or energy expansion, and to focus directly on them simplifying the calculation. Moreover, in the case of an high-energy expansion, manifest power-counting would permit to expand separately the amplitude of each individual diagram, allowing for instance to neglect the masses in the internal line propagators, which is often a substantial simplification. In the ordinary covariant formulation this is not possible, the positive powers of $E/m$ from the polarization vectors cancel negative powers from the Taylor expansion of the propagators, so that the masses have to be retained until the end of the calculation. Our formulation of massive gauge theories, which we call ``Equivalent Gauge'', overcomes the above-mentioned issues.

A partial solution to the problem of bad energy behavior is offered by a very well-known result, the so-called ``Equivalence Theorem''. In its ``strong'' formulation \cite{Chanowitz:1985hj} (see \cite{Horejsi:1995jj} for a review), the theorem states that the longitudinally polarized vectors are equivalent, in the high-energy limit, to the corresponding scalar Goldstone bosons.\footnote{Beyond tree-level, this only holds up to multiplicative corrections \cite{Bagger:1989fc}, in the following we will discuss this point in detail.} The high energy amplitudes involving longitudinal particles can thus be computed from Goldstone diagrams, without having to deal with the badly-behaved polarization vectors. Inspired by the Equivalence Theorem, the central idea of the present work is that in order to obtain well-behaved polarization vectors one should manage to change the way in which we represent of the $W_L$ particle as a state in the Fock space of the theory, in a way that it assumes a component along the excitations of the Goldstone field. From the Equivalence Theorem we expect that, provided the shift is performed in the proper way, the Goldstone component will dominate at high energy while the component along the gauge field will vanish and the bad energy behavior problem will be avoided. Changing the representation of the $W_L$ is not difficult at all, if one exploits the BRS invariance of the theory. Actually, we should remember that the definition of physical states in a gauge theory is conventional anyhow, the physical states are the elements of the BRS cohomology, and there are infinitely many equivalent ways to represent them in terms of the states in the Fock space. In particular, a valid representative is obtained from the standard one by performing a shift with a BRS-exact state. By such a shift we will make the $W_L$ assume a component along the Goldstone and we will complete the programme previously outlined.

The ambiguity in the definition of the states is due to gauge invariance, fixing it is therefore part of the gauge-fixing procedure. Our formulation of the theory, where the $W_L$ are represented in a non-standard way, is in this sense a ``gauge choice''. However the Equivalent Gauge is not a new gauge, in the sense that it does not involve new exotic gauge-fixing conditions or gauge-fixing functionals. It is formulated as an ordinary $R_\xi$ gauge and therefore its Feynman rules for vertices and propagators are completely standard. What changes is only the Feynman rule for external longitudinal particles.

The paper is organized as follows. In Section~\ref{s11} we set up our conventions and describe the specific model where, for definiteness, the formulation of the Equivalent Gauge will be discussed. In Section~\ref{freeth} we illustrate the Equivalent Gauge in the simple case of the free theory, while the treatment of the complete interacting theory is postponed to Section~\ref{it}. As it will become clear in the following, the derivations of Section~\ref{it} rely on certain gauge-fixing conditions, which we make explicit in Section~\ref{mf}. Additional technical details, related with the LSZ reduction formula and with Slavnov-Taylor identities, are reported in Appendix~\ref{RF} and Appendix~\ref{ST}, respectively. Finally, we present our conclusion in Sect.~\ref{conc}.

\section{The Equivalent Gauge}

\subsection{The Model}\label{s11}

In oder to keep the discussion simpler, we will illustrate the Equivalent Gauge in the context of the so-called Higgs-Kibble model, which is an \mbox{SU$(2)_L$} gauge theory with one scalar Higgs doublet that takes a VEV and breaks \mbox{SU$(2)_L$} completely. The inclusion of the Hypercharge group of the SM would require additional work, and will not be discussed here.

Before gauge-fixing, the Lagrangian reads
\beq
\displaystyle
{\mc L}_0\,=\,-\frac12\,\tr\left[W_{\mu\nu}W^{\mu\nu}\right]\,+\,
\tr\left[\left(D_{\mu}\mc{H}\right)^\dagger D^{\mu}\mc{H}\right]\,-\,
\frac{\lambda}{4}\left\{\tr\left[\mc{H}^\dagger\mc{H}\right]-\frac{{\tilde{v}}^2}2\right\}^2\,,
\label{lag00}
\eeq
where we represented the Higgs field by a $2\times2$ pseudo-real matrix $\mc{H}$, defined as
\beq
\displaystyle
\mc{H}\,=\,\frac{v+h}2\,\I\,+\,i\,\frac12 \sigma^a \pi_a\,,
\eeq
in terms of the three Pauli matrices $\sigma^a$. The four real fields $h$ and $\pi_a$ correspond, respectively, to the physical Higgs particle and the three Goldstone bosons. In the above equation we denote as ``$v$'' the VEV of the Higgs field, which of course differs, beyond the tree-level order, from the VEV parameter $\t{v}$ which appears in the Lagrangian. For completeness we report our definition of the gauge field strength and of the Higgs covariant derivative
\beq
\displaystyle
W_{\mu\nu}=\partial_\mu W_\nu-\partial_\nu W_\mu-i\,g\left[W_\mu,W_\nu\right]\,,\;\;\;\;\;
D_\mu\mc{H}\,\equiv\,\partial_\mu\mc{H}\,-\,i\,g\,W_\mu\mc{H}\,.
\eeq
The gauge connection $W_\mu$ is expanded as $W_\mu\equiv W_\mu^a\sigma_a/2$ in terms of the three canonically normalized gauge fields $W_\mu^a$. In our notation, the  \mbox{SU$(2)_L$} gauge transformation acts on $\mc{H}$ as multiplication from the left, $\mc{H}(x)\rightarrow\Omega_2(x) \mc{H}(x)$.

On top of gauge symmetry, the Higgs-Kibble Lagrangian is invariant under the global custodial group \mbox{SO$(3)_c$}, and the advantage of the Higgs matrix notation is that it makes this manifest. Under \mbox{$\gamma\in\,$\mbox{SO$(3)_c$}}, the fields transform as
\beq
\displaystyle
\mc{H}\,\rightarrow\, \gamma \mc{H}\gamma^\dagger\,,
\;\;\;\;\;
W_\mu\,\rightarrow\,\gamma W_\mu \gamma^\dagger\,,
\eeq
so that the three Goldstone bosons and the gauge fields are custodial triplets, while the Higgs is a scalar. Differently from $\mbox{SU$(2)_L$}$, the custodial group is not broken by the Higgs VEV. Our derivations will make frequent use of this unbroken symmetry, for this reason it would not be completely straightforward to generalize them to the SM where custodial symmetry is broken by the gauging of hypercharge. 

With the standard Faddeev-Popov method, the theory can be reformulated by introducing a gauge-fixing term, the ghost fields $\o$ and the anti-ghosts $\ob$. We adopt the canonical Feynman-'t Hooft $R_\xi$ gauge-fixing functional
\beq
\displaystyle
f_a\,\equiv\,\partial_\mu W^{\mu}_a\,+\,\widetilde{m}\,\xi \pi_a\,,
\label{gff}
\eeq
and the gauge-fixed Lagrangian reads
\beq
\displaystyle
{\mc L}\,=\,{\mc L}_0\,-\,\frac1{2\xi}\left(f_a\right)^2\,+\,{\mc L}_{gh}\,,
\label{gfl}
\eeq
with the ghost/anti-ghost term given by
\beq
\displaystyle
{\mc L}_{gh}\,=\,
-\ob_a\partial_\mu\left(\partial^\mu \o^a
+g\,\varepsilon^{abc}W^\mu_b\o_c\right)
-\frac{1}{2}g\,\widetilde{m}\,\xi\,\ob_a \left[(v+h)\omega^a
+\varepsilon^{abc}\pi_b\o_c
\right]
\,.
\label{lgh}
\eeq
Notice that in choosing the gauge-fixing functional we have been careful not to break the custodial group, so that the gauge-fixed Lagrangian is still invariant under \mbox{SO$(3)_c$}, with $\o$ and $\ob$ transforming as triplets.

Differently from the original one, the gauge-fixed theory is suited to set up a perturbative expansion in terms of Feynman diagrams, through which it will be possible to compute the correlation functions and eventually the scattering amplitudes. Of course, the gauge-fixed theory is  unphysical by itself, and indeed it depends on two unphysical gauge-fixing parameters $\widetilde{m}$ and $\xi$. Only some observables are physical, namely the scattering amplitudes among physical particles, and are independent of $\widetilde{m}$ and $\xi$. While the final physical results will be independent of the gauge-fixing parameters, the intermediate steps of the calculations do depend on their value, so that choosing them in a given way might be more or less convenient for certain applications. Fixing the parameters in a convenient way is precisely what is called ``choosing a gauge'' in the common terminology, we will illustrate our choice in Section~\ref{mf}. 

However there is one further ambiguity besides the choice of $\widetilde{m}$ and $\xi$, which is also related with the gauge invariance of the original theory. This is the fact that the space of physical states is not embedded in a unique way in the extended Fock space of the gauge-fixed theory. The physical Hilbert space is defined as the cohomology of the BRST charge $Q$ {\it{i.e.}}, poorly speaking, as the states which are close but not exact under $Q$. With this definition, the physical states can be represented in various ways in terms of the unphysical ones. Namely, it is always possible to add a BRS exact state, of the form $Q|\psi\rangle$, to the definition of the physical particles. The choice of the physical states, very much like the choice of $\widetilde{m}$ and $\xi$, is part of the gauge-fixing procedure. The physical scattering amplitudes are of course completely insensitive to the choice of the states, but the Feynman rules of the gauge-fixed theory, through which these amplitudes are computed in perturbation theory, do depend on it. In particular, the wave function factors associated with external physical particles will be affected. The essence of the Equivalent Gauge is to represent the longitudinally polarized vector boson in a way that its wave function does not suffer of an anomalous high-energy behavior, namely it does not grow with the energy.

\renewcommand{\arraystretch}{1.5}
\setlength{\tabcolsep}{3.5pt}
\begin{table}
\begin{center}
\begin{tabular}{rl}
$s(W_\mu^a)\,=$&$\left[i\,Q, W_\mu^a\right]\,=\, \partial^\mu \o^a
+g\,\varepsilon^{abc}W^\mu_b\o_c$\\
$s(\pi^a)\,=$&$\displaystyle\left[i\,Q, \pi^a\right]\,=\frac{g}2\left(v+h\right)\o^a+\frac{g}2\,\varepsilon^{abc}\pi_b
\o_c$\\
$s(h)\,=$&$\displaystyle\left[i\,Q, h\right] \, =\, -\frac{g}2\o_a\pi^a$\\
$s(B^a)\,=$&$\left[i\,Q, B^a\right]\,=\,0$\\
$s(\ob_a)\,=$&$\left\{i\,Q, \ob_a\right\}\,=\, B_a$\\
$s(\o_a)\,=$&$\displaystyle\left\{i\,Q, \o_a\right\}\,=\,-\frac12 g\,\varepsilon_{abc} \o^b\o^c$
\end{tabular}
\end{center}
\caption{The BRS variation $s(\Phi)$ for each of our fields. The (anti-)commutators of the BRS charge $Q$ with the field operators are defined according to eq.~(\ref{brsch}).}
\label{table:brs}
\end{table}

In order to identify the physical states we must further rewrite our theory in a way that makes manifest its invariance under BRS transformations. We thus introduce a triplet of scalar auxiliary fields $B^a$ and write the Lagrangian as
\beq
{\mc{L}}={\mc{L}}_0+\frac\xi2 B^aB_a+B^af_a\,+{\mc{L}}_{gh}\,.
\label{lagtot0}
\eeq
The latter is completely equivalent to eq.~(\ref{gfl}), as one can easily check by substituting the equations of motion of the auxiliary field 
\beq
\displaystyle
\xi\,B_a\,=\,-\,f_a\,.
\label{auxeom}
\eeq
It is a standard textbook exercise to verify that the Lagrangian (\ref{lagtot0}) is invariant under BRS transformations
\beq
\displaystyle
\Phi\;\;\rightarrow\;\;\Phi\,+\,\epsilon\, s\left(\Phi\right)\,,
\eeq
where the infinitesimal parameter $\epsilon$ is taken to commute with the bosonic fields and to anti-commute with the fermionic ones. The BRS variations $s(\Phi)$ are reported in Table~\ref{table:brs}, notice that the variations of the $W$, $\pi$ and $h$ fields correspond to an infinitesimal gauge transformation with parameter $\epsilon\,\o$. In the operator language, the BRS transformations are generated by the charge $Q$, whose action is defined as
\beq
\left[\epsilon\,i\,Q,\,\Phi\right]\equiv\epsilon s\,\left(\Phi\right)\,.
\label{brsch}
\eeq
Since $\epsilon$ commutes with bosons and anti-commutes with fermions, the above equation leads, respectively, to commutation and anti-commutation relations as in Table~\ref{table:brs}.

Several interesting conclusions can be reached by looking at Table~\ref{table:brs}. First of all, one can check that the BRS transformation squares to zero, {\it{i.e.}} $s\left(s(\Phi)\right)=0$, which means the BRS charge is a nilpotent operator
\beq
Q^2\,=\,0\,.
\label{nil}
\eeq
Moreover, following Ref.~\cite{KO}, we notice that $Q$ can be Hermitian
\beq
Q^\dagger\,=\,Q\,,
\label{her}
\eeq
only if the ghost and anti-ghost fields are, respectively, Hermitian and anti-Hermitian operators
\beq
\o^\dagger\,=\,\o\,,\;\;\;\;\;\ob^\dagger\,=\,-\,\ob\,.
\label{oher}
\eeq
Because of their anti-commutative nature, the ghosts and anti-ghosts having opposite Hermiticity is precisely what is needed to make their Lagrangian (\ref{lgh}) real. Finally, we see from Table~\ref{table:brs} that $Q$ is a Lorentz scalar. Thus it commutes with the Lorentz generators and also, since it is conserved, with the $4$-momentum
\beq
\left[Q,\,J^{\mu\nu}\right]\,=\,0\,,\;\;\;\;\;\left[Q,\,P^\mu\right]\,=\,0\,.
\label{comm}
\eeq
After performing the canonical quantization of the gauge-fixed Lagrangian, one could derive the BRS charge operator and verify explicitly all the properties listed in eq.s~(\ref{nil}), (\ref{her}) and (\ref{comm}). This derivation will not be needed for our purposes and thus it will not be repeated here, the reader is referred to the original literature \cite{KO}.\footnote{Actually Ref.~\cite{KO} only considered covariant $\xi$ gauges, which corresponds to the particular case $\widetilde{m}=0$ in eq.~(\ref{gff}). However the results can be straightforwardly generalized.}

On top of custodial and BRS, our theory has two more exact symmetries we will make use of. The first one is ghost number which, again following \cite{KO}, acts on the ghosts and to the anti-ghosts as
\beq
\o\,\rightarrow e^\lambda\o\,,\;\;\;\;\;\ob\,\rightarrow e^{-\lambda}\ob\,,
\eeq
while leaving all other fields invariant. We typically rephrase the above equation by saying that $\o$ and $\ob$ have, respectively, ghost number equal to $+1$ and to $-1$. Notice however that the parameter $\lambda$ has to be real in order to preserve the Hermiticity properties of the fields, therefore ghost number acts as a rescaling rather than a \mbox{U$(1)$} transformation. Thus we should be talking of the ghost ``weight'' rather the ghost ``number'', nevertheless we will keep using this improper, but conventional, terminology. For what concerns the BRS charge, the commutation relations of Table~\ref{table:brs} tell us that it has ghost number equal to $+1$.

The last symmetry to be discussed is CPT, under which our gauge-fixed theory is invariant like any other relativistic local quantum field theory.\footnote{Our theory is also separately invariant under C, P and T, but this will not enter in our discussion.} The CPT operator, denoted as $\Theta$ for shortness, acts as follows
\beq
\begin{tabular}{lcl}
$W_\mu(x)\,\rightarrow\,\Theta W_\mu(x) \Theta^{-1}\,=\,-\,W_\mu(-x)\,,$&\;\;\;\;\;&$\pi(x)\,\rightarrow\,\Theta \pi(x) \Theta^{-1}\,=\,\pi(-x)\,,$\\
$h(x)\,\rightarrow\,\Theta h(x) \Theta^{-1}\,=\,h(-x)\,,$&\;\;\;\;\;&$B(x)\,\rightarrow\,\Theta B(x) \Theta^{-1}\,=\,B(-x)\,,$\\
$\o(x)\,\rightarrow\,\Theta \o(x) \Theta^{-1}\,=\,\o(-x)\,,$&\;\;\;\;\;&$\ob(x)\,\rightarrow\,\Theta \ob(x) \Theta^{-1}\,=\,\ob(-x)\,.$
\end{tabular}
\label{cptf}
\eeq
The only terms in the above equation that require some comment are those in the last line. In spite of having opposite Hermiticity, $\o$ and $\ob$ transform in the same way under CPT, which means that the $\ob$ field has an intrinsic phase equal to $-1$ with respect to the canonical action of CPT on the scalars, which would be $\phi(x)\rightarrow\phi^\dagger(-x)$. This minus sign is essential to make the Lagrangian in eq.~(\ref{lagtot0}) transform as ${\mathcal{L}}(x)\rightarrow{\mathcal{L}}(-x)$, leaving the action invariant. Finally, again looking at Table~\ref{table:brs} we can see how the BRS charge transforms under CPT. Not surprising, $Q$ is odd like any other internal symmetry generator, {\it{i.e.}}
\beq
\Theta\,Q\,\Theta^{-1}\,=\,-\,Q\,.
\label{cptq}
\eeq

\subsection{Free Theory}\label{freeth}

Many structural features of the model and the essence of the Equivalent Gauge can be illustrated in the free limit. This is the aim of the present section, we will discuss in the following one how to take care of interactions. We are thus going to consider the free theory, and moreover we are going to make a particularly simple choice for the gauge-fixing parameters $\widetilde{m}$ and $\xi$, namely
\beq
\widetilde{m}\,=\,m\,,\;\;\;\;\;\xi\,=\,1\,,
\label{gch}
\eeq
where $m$ is the pole mass of the physical $W$ bosons, which in the free case reads
\beq
\displaystyle
m\,=\,\frac12\, g\,v\,.
\label{freem}
\eeq
In the free limit the Lagrangian (\ref{gfl}) simplifies dramatically and becomes
\bea
{\mc{L}}_{\textrm{free}}&=
&-\frac14\partial_\mu W_\nu \partial^\mu W^\nu
+\frac{m^2}2W_\mu W^\mu\nonumber\\
& \ &+\frac12\partial_\mu\pi\partial^\mu\pi -\frac{m^2}2\pi^2 \nonumber\\
& \ &
+\,\partial^\mu\ob\,\partial_\mu\o-m^2\,\ob\,\o\,,
\label{lagfree}
\eea
where we omitted, for shortness, the custodial triplet indices ``$a$'' and the terms involving the Higgs field, which just describe the physical Higgs particle and will not play any role in what follows.

The gauge-fixing choice of eq.~(\ref{gch}) makes the free theory extremely easy to quantize in the canonical formalism.\footnote{For generic gauge-fixing parameters canonical quantization is more involved, see Ref.s~\cite{KO,Nakanishi:1972sm}.} The first condition, $\mt=m$, cancels the \mbox{$W$-$\pi$} mixing and renders the equations of motion of second order in derivatives. This equivalently means that the momentum-space propagators only contain single-pole singularities contrary to the general case $\mt\neq m$ where double poles, the so-called ``dipole terms", do appear, obscuring the interpretation of the theory in terms of propagating relativistic particles. With the second condition, $\xi=1$, the equations of motion further simplify and become just $(\Box +m^2)\Phi=0$, with the same mass $m$ of eq.~(\ref{freem}) for all the fields. Therefore all the particles of our theory, both the physical and the unphysical ones, will share a common pole mass. The degeneracy of the spectrum and the absence of dipole terms in the propagators are essential ingredients for the formulation of the Equivalent Gauge, we will show in Section~\ref{mf} how these conditions can be imposed also in the interacting theory by a suitable choice of the gauge-fixing parameters.

Upon quantization, the field operators are expanded as
\bea
\displaystyle
W_\mu(x)=&&\hspace{-16pt}\int\hspace{-4pt}\frac{d^3p}{(2\pi)^3}\frac{e^{-ipx}}{\sqrt{2E_{\p}}}
\left[
\sum_{h=\pm,0}\epsilon_\mu^h(\p)\widetilde{w}_h(\p)
+\epsilon_\mu^s(\p)\widetilde{s}(\p)
\right]
\,+\,\textrm{h.c.}\,.\nn\\
\pi(x)=&&\hspace{-16pt}\int\hspace{-4pt}\frac{d^3p}{(2\pi)^3}\frac{e^{-ipx}}{\sqrt{2E_{\p}}}
\,\widetilde{g}(\p)
\,+\,\textrm{h.c.}\,,\nn\\
\o(x)=&&\hspace{-16pt}\int\hspace{-4pt}\frac{d^3p}{(2\pi)^3}\frac{e^{-ipx}}{\sqrt{2E_{\p}}}
\,\ot(\p)
\,+\,\textrm{h.c.}\,,\nn\\
\ob(x)=&&\hspace{-16pt}\int\hspace{-4pt}\frac{d^3p}{(2\pi)^3}\frac{e^{-ipx}}{\sqrt{2E_{\p}}}
\,\obt(\p)
\,-\,\textrm{h.c.}\,,
\label{mdec}
\eea
where, because of the mass degeneracy, the same basis of plane waves, with frequency \mbox{$p_0=E_{\p}=\sqrt{\p^2+m^2}$}, is used for the decomposition of all the fields. Notice that because of the Hermiticity properties of eq.~(\ref{oher}) only two set of independent creation/annihilation operators, rather than four, are present in the ghost sector. Correspondingly, we will have only two particles, $|\o\rangle$ and $|\ob\rangle$, with fermionic statistic and non-vanishing ghost number. The gauge field polarization vectors are 
\bea
&&\epsilon_\mu^\pm(\p)=\left\{0,\,{\vec{\epsilon}}^\pm(\p)\right\}\,,\nn\\
&&\epsilon_\mu^0(\p)=\frac1m\left\{|\p|,\,-\frac{E_\p}{|\p|}\p\right\}\,,\nn\\
&&\epsilon_\mu^s(\p)=\frac{i}mp_\mu=\frac{i}m\left\{E_\p,\,-\p\right\}\,,
\label{pvec}
\eea
and verify the standard normalization and completeness relations
\beq
\left(\epsilon_\mu^r\right)^*\eta^{\mu\nu}\epsilon_\nu^{r'}=-\zeta_r\delta_{rr'}
\,,\;\;\;\;\;
\sum_{r=\pm,0,s}\zeta_r\left(\epsilon_\mu^r\right)^*\epsilon_\nu^{r}=-\eta_{\mu\nu}\,,
\label{crel}
\eeq
with $\zeta_{\pm,0}=+1$ and $\zeta_{s}=-1$. 

\renewcommand{\arraystretch}{1.5}
\setlength{\tabcolsep}{3.5pt}
\begin{table}
\begin{center}
\begin{tabular}{rccccccl}
\ & $w_{h}$ &  $s$ & $g$ & $\o$ & $\ob$ \\
\multirow{5}{*}{$\mathcal{N}\,=\left( \vphantom{\begin{array}{c}
 +1 \\ +1\\ +1 \\ +1 \\ +1 
\end{array}} \right.\hspace{-4pt}$} 
& $\delta_{h\,h'}$ & $0$ & $0$ & $0$ &$0$ & \multirow{5}{*}{$
\hspace{-4pt}\left. \vphantom{\begin{array}{c}
 +1 \\ +1\\ +1 \\ +1 \\ +1 
\end{array}} \right)\hspace{-4pt}
\begin{array}{c}
w_{h'}\\  s\\ g \\\o\\\ob
\end{array}$}\\
& $0$ & $-1$ & $0$ & $0$ & $0$  & \\
& $0$ & $0$ & $+1$ & $0$ &$0$  & \\
& $0$ & $0$ & $0$ & $0$ & $-1$  & \\
& $0$ & $0$ & $0$ & $-1$ & $0$  & \\
\end{tabular}
\end{center}
\caption{The norm matrix in the single-particle subspace.}
\label{table:norm}
\end{table}

The creation/annihilation operators obey commutation/anti-commutation relations
\bea
\displaystyle
&\left[\wt_h(\p),\wt_{h'}^\dagger(\q)\right]=(2\pi)^3\delta_{hh'}\delta^3(\p\hspace{-2pt}-\hspace{-2pt}\q)\,,&\nn\\
\displaystyle
&\Big[\widetilde{g}(\p),\widetilde{g}^\dagger(\q)\Big]=(2\pi)^3\delta^3(\p\hspace{-2pt}-\hspace{-2pt}\q)\,,&\nn\\
\displaystyle
&\Big[\widetilde{s}(\p),\widetilde{s}^\dagger(\q)\Big]=-(2\pi)^3\delta^3(\p\hspace{-2pt}-\hspace{-2pt}\q)\,,&\nn\\
\displaystyle
&\Big\{\ot(\p),\obt^\dagger(\q)\Big\}=\Big\{\obt(\p),\ot^\dagger(\q)\Big\}=-(2\pi)^3\delta(\p\hspace{-2pt}-\hspace{-2pt}\q)\,.&
\label{ca}
\eea
All the other commutators, and the anti-commutators among fermionic operators, vanish. Each creation operator acting on the vacuum defines a single-particle state
\beq
|\psi_I\rangle=\sqrt{2 E_{\p}}\,\widetilde\psi_I^\dagger\,|0\rangle\,,
\eeq
whose norm is immediately computed through the canonical commutators of eq.~(\ref{ca})
\beq
\langle\psi_I(\p)|\psi_J(\q)\rangle= 2 E_{\p} (2\pi)^3\delta^3(\p-\q)\times {\mathcal{N}}_{IJ}\,,
\label{nor}
\eeq
with the matrix ${\mathcal{N}}$ reported in Table~(\ref{table:norm}). Our theory describes one Lorentz triplet $|w_h\rangle$ of massive spin one states, with helicity $h=\pm1,0$ and positive norm, plus four Lorentz scalars $|s\rangle$, $|g\rangle$, $|\o\rangle$ and $|\ob\rangle$. The scalar polarization state $|s\rangle$ and the Goldstone $|g\rangle$ have, respectively, negative and positive norm, while the ghost and the anti-ghost have an off-diagonal norm matrix. The norm in the ghost sector could be diagonalized by a change of basis, leading to one positive and one negative norm state. However our basis is more convenient because $|\o\rangle$ and $|\ob\rangle$ have definite ghost number, equal to $+1$ and to $-1$, respectively.

From the mode decomposition of eq.~(\ref{mdec}) we can also work out, for future use, the action of the CPT operator on the single-particle states. From the definition in eq.~(\ref{cptf}) we find that CPT acts in the canonical way, without extra phases, on all particles, {\it{i.e.}}
\beq
\Theta |w_h(\p)\rangle\,=\,-\,(-)^h|w_{-h}(\p)\rangle\,,\;\;\;\;\;\Theta |S_i(\p)\rangle\,=\,|S_i(\p)\rangle\,,
\label{cptp}
\eeq
where we denoted as $|S_i\rangle$, for shortness, all the scalar states of the theory. Notice that the above result relies on the unconventional imaginary factor in the definition (\ref{pvec}) of the scalar polarization vector $\epsilon_\mu^s$.

Needless to say, not all the particles are physical, and indeed the norm matrix is not positive-definite. The physical states, with positive norm, are represented by the cohomology of the BRS operator $Q$. In practice, this means the physical states live in the kernel of $Q$, and that two physical states are regarded as independent only if their difference is not a BRS-exact state, of the form $Q|\psi\rangle$. The action of $Q$ is given in Table~\ref{table:brs} for the case of the interacting theory, the free limit is taken by dropping all the terms which are quadratic in the fields. By using the equation of motion of the auxiliary field $B$ in eq.~(\ref{auxeom}) and substituting the mode decomposition of eq.~(\ref{mdec}) we can turn Table~\ref{table:brs} into a set of commutators (and anti-commutators, for fermions) of $Q$ with the creation/annihilation operators, 
\bea
& \left[Q, \wt_{h}(\p)\right]\,=\,\left\{Q, \ot(\p)\right\}\,=\,0\,,\;\;\;\;\; & \left[Q, \widetilde{s}(\p)\right]\,=\, i\,m\,\ot(\p)\,,\nn\\
& \left[Q, \widetilde{g}(\p)\right]\,=\, -i\,m\,\ot(\p)\,,\;\;\;\;\;\;\;\;& \{Q, \obt(\p)\}\,=\,i\,m\,\left[\widetilde{s}(\p)+\widetilde{g}(\p)\right]\,.
\label{brsmode}
\eea
From the equation above we immediately derive the action of $Q$ on the states
\bea
& Q|w_h\rangle\,=\,Q|\o\rangle\,=\,0\,,\;\;\;\;\; & Q|s\rangle\,=\,i\,m\,|\o\rangle\,,\nn\\
& Q|g\rangle\,=\,-i\,m\,|\o\rangle\,, \;\;\;\;\;\;\;\;& Q|\ob\rangle\,=\,-i\,m\,\left(|s\rangle+|g\rangle\right)\,.
\label{brsfree}
\eea

We can now characterize the BRS cohomology, and discuss the various ways in which it can be represented in terms of the particles in the extended Fock space of the theory. The standard approach is to take the three $|w_h\rangle$'s as the representatives of the cohomology and to consider all other states as unphysical. This is definitely a consistent way to proceed because the \mbox{$|w_h\rangle$'s} belong to the kernel of $Q$ and also, differently from the $|\o\rangle$ state, they are not BRS-exact. With this choice, the three physical polarizations of the massive $W$ boson are described by the \mbox{$|w_h\rangle$'s}, and in particular the longitudinal polarization is given by
\beq
|W_L\rangle_{\textrm{standard}}\,=\,|w_0\rangle\,.
\label{stdL}
\eeq

However this standard choice is not unique, one can construct an entire family of equally valid representatives by adding BRS-exact states to the standard definition. The freedom of picking up one definition or the other is associated with the gauge invariance of the original theory, and it can be intuitively understood as the freedom of performing a gauge transformation of the fields which describe the external particles. Choosing the representative of the physical state is part of the gauge-fixing procedure, very much like choosing the gauge-fixing parameters $\widetilde{m}$ and $\xi$. The essential idea of the Equivalent Gauge is to modify the longitudinal $W$ representative, with respect to the standard one, by adding one BRS-exact state with vanishing ghost number. Namely, we define
\beq
\displaystyle
|W_L\rangle=|w_0\rangle+\frac1{m}\,Q\,|\ob\rangle=|w_0\rangle-i|s\rangle-i|g\rangle\,,
\label{egL}
\eeq
while we maintain the standard definition for the transverse polarizations. The physical longitudinal particle is now represented as the sum of $|w_0\rangle$, of the scalar and of the Goldstone states. Notice that $|W_L\rangle$, as defined above, is perfectly physical, and indeed it has positive norm exactly like the ``standard'' longitudinal state $|w_0\rangle$. Intuitively, the reason for this definition is that the standard longitudinal polarization vector $\epsilon_\mu^0$, associated with the standard longitudinal state $|w_0\rangle$, diverges like \mbox{$\epsilon_\mu^0\to p_\mu/m$} in the high energy limit. By subtracting $|s\rangle$, with polarization \mbox{$\epsilon_\mu^s = i\,p_\mu/m$}, we will cancel the divergence and obtain a well-behaved polarization vector.

In order to see how this works in detail, let us compute the Feynman rule associated with an external longitudinal $W$, and discuss how it changes when we switch from the standard definition of the state (\ref{stdL}) to the one of the Equivalent Gauge in eq.~(\ref{egL}). Obviously this change will not affect the final result provided we compute physical quantities, {\it{i.e.}} the matrix elements of gauge-invariant operators. To derive the rule, let us write down the matrix elements of the fields among the vacuum and the single particle states. Focusing on the bosonic sector we have
\beq
\begin{array}{l}
\langle0|W_\mu(x)|w_h(\p)\rangle=\epsilon_\mu^h(\p)\,e^{-ipx}\,,\\
\langle0|W_\mu(x)|s(\p)\rangle=-\,\epsilon_\mu^s(\p)\,e^{-ipx}\,,\\
\langle0|\pi(x)|g(\p)\rangle=e^{-ipx}\,,
\end{array}
\label{renf}
\eeq
where the negative sign in the scalar state matrix element is due to its negative norm.
\begin{figure}
\begin{center}
\includegraphics[width=1\textwidth]{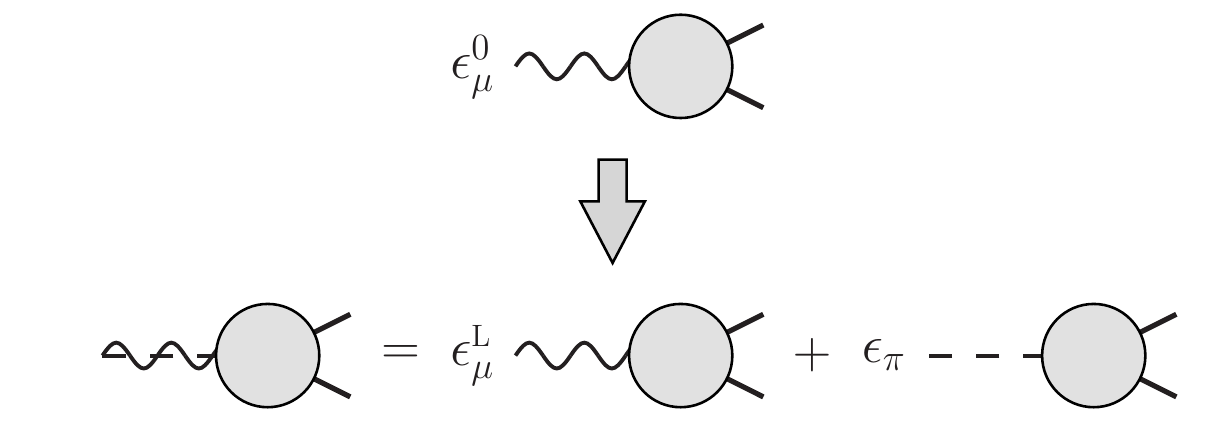}
\end{center}
\caption{Feynman rules for longitudinally polarized incoming $W$'s. The standard rule is depicted on the upper part, while the lower one shows how it gets modified in the Equivalent Gauge.}
\label{frule}
\end{figure}
Now, imagine computing the matrix element of some time-ordered product of fields with one $|W_L\rangle$ as incoming external particle. With the standard definition of eq.~(\ref{stdL}), the incoming $|W_L\rangle$ can be annihilated only by the action of the $W_\mu$ field operator, and therefore its Feynman rule is depicted as in the upper part of Figure~\ref{frule}, with one external gauge field line entering into the diagram. When $W_\mu$ annihilates the state, it leaves behind, in the momentum space, a wave-function factor 
\beq
\epsilon_\mu^0(\p)=\frac1m\left\{|\p|,\,-\frac{E_\p}{|\p|}\p\right\}\,.
\label{stdpol}
\eeq
Instead, consider the Equivalent Gauge definition of $|W_L\rangle$ in eq.~(\ref{egL}). In this case the incoming state can be annihilated by two different fields. Either by $W_\mu$, which can annihilate $|w_0\rangle$ or $|s\rangle$, or by the Goldstone boson field $\pi$, which annihilates the Goldstone state $|g\rangle$. The corresponding matrix elements are
\beq
\begin{array}{l}
\langle0|W_\mu(x)|W_L(\p)\rangle=\langle0|W_\mu(x)\cdot\big[|w_0(\p)\rangle
-i\,|s(\p)\rangle\big]=\left[\epsilon_\mu^0(\p)+i\,\epsilon_\mu^s(\p)\right]\,e^{-ipx}\,,\\
\langle0|\pi(x)|W_L(\p)\rangle=-i\,\langle0|\pi(x)|g(\p)\rangle=-i\,e^{-ipx}\,.
\end{array}
\eeq
Therefore, in the Equivalent Gauge, the matrix element of longitudinal $W$'s will receive two contributions, one from the diagrams with one external $W_\mu$, and one from those with external $\pi$. The wave function factors for these two class of diagrams are, respectively
\bea
&&\epsilon_\mu^L(\p)\equiv \epsilon_\mu^0(\p)+i\,\epsilon_\mu^s(\p)=-\frac{m}{E_\p+|\p|}
\left\{1,\,\frac{\p}{|\p|}\right\}\,,\nn\\
&&\epsilon_\pi(\p)\equiv-i\,.
\label{freg}
\eea
The situation is well represented by a double line notation as in Figure~\ref{frule}, a similar notation was proposed also in Ref.~\cite{Veltman:1989ud}. 

As illustrated by the picture, the double line means that two sets of Feynman diagrams need to be drawn for each external $|W_L\rangle$, one with a gauge and the other with a Goldstone boson external line. This proliferation of diagrams is a complication at the practical level, but not such a serious one because computing some more diagram is not a big issue with the powerful automated tools to our disposal. The advantage is that the wave-function factors associated with the Feynman rule are now well-behaved with the energy, differently from the ``standard'' longitudinal polarization vector in eq.~(\ref{stdpol}) which grows like $E/m$. As discussed at length in the Introduction, this growth is problematic because it obscures the energy behavior of the amplitudes, introducing extra powers of $E/m$ which often cancel in the final result from a complicated conspiracy of different diagrams. Similarly, the $E/m$ terms also obscures the power-counting of the gauge coupling $g$. Since $m=gv/2$, negative powers of $m$ are negative powers of $g$, which cancel positive powers from the Feynman vertices. It might instead be useful to have the $E$ and $g$ power-counting under control, this would allow to select the most relevant diagrams for a given process, and to simplify the calculation under certain approximations. The Equivalent Gauge makes power-counting manifest. Indeed we see in eq.~(\ref{freg}) that, thanks to the judicious choice of the factor $1/m$ in eq.~(\ref{egL}), the growth with the energy of the longitudinal polarization vector is exactly canceled. 

But there is more than that. The new polarization vector $\epsilon_\mu^L$ not only does not grow, but it vanishes as $m/E$ at large energy, while the Goldstone wave function term $\epsilon_\pi$ stays constant, equal to $-i$. Therefore the contribution of the diagrams with external gauge fields will be suppressed in the high energy limit, and the ones with the Goldstones will become relatively more important. This result is nothing but the Equivalence Theorem \cite{Chanowitz:1985hj}, which states that in the high energy limit the amplitudes involving longitudinal $W$'s reduce, up to a phase, to the ones of the associated Goldstone bosons.\footnote{There is an interesting caveat in the above argument. Even if there is a power-like suppression of the gauge contribution coming from the polarization vector, in order to conclude that the Goldstone one dominates one has to assume that this suppression is not compensated by a different energy scaling of the Goldstone and of the gauge Feynman amplitudes. However there are plenty of cases where the Goldstone diagrams receive an additional $m/E$ power suppression because of selection rules \cite{Borel:2012by,Coradeschi:2012iu}. For instance this occurs for some specific polarized $WW\to WW$ scattering amplitudes, and indeed the Equivalence Theorem is violated in these examples. Of course this is not an issue in the Equivalent Gauge, it simply means that both the gauge and the Goldstone contributions will have to be retained, as they scale with energy in the same way. Furthermore, exactly because of the suppression, all these amplitudes will typically give a negligible contribution to physical processes, so that we will most likely never need to compute them in practice.\label{foot}} The Equivalent Gauge makes transparent the physical origin of the Equivalence Theorem, and furthermore, since it is an exact reformulation of the theory, allows one to estimate or compute the $m/E$ corrections to the Equivalence Theorem limit in a systematic way. 

This discussion almost concludes our illustration of the Equivalent Gauge in the free theory. There is only one last point, which is rather trivial in the free case but worth discussing in view of the generalization to the interacting one. This is the fact that until now we have only considered the states of a single isolated particle, while obviously we will have to deal with multi-particle states in order to compute the scattering amplitudes. Our new definition of a single longitudinal particle in eq.~(\ref{egL}) is definitely a consistent one, because it is equivalent to the standard choice of eq.~(\ref{stdL}), but one might wonder if this equivalence survives when the longitudinal is part of a multi-particle state. This is actually the case, and can be verified by noticing that the BRS charge acts as the tensor product representation on multi-particle states, {\it{i.e.}}
\beq
Q\,\left\{|\Psi_{1}\rangle\ldots|\Psi_{n}\rangle\right\}\,=\,\sum_{k=1,n}(-)^{p_k}
|\Psi_{1}\rangle\ldots\left\{Q\,|\Psi_{k}\rangle\right\}\ldots |\Psi_{k}\rangle\,.
\label{Qtens}
\eeq
The above result is immediately derived by repeatedly applying eq.~(\ref{brsmode}). Notice the presence of the factor $(-)^{p_k}$, which counts the number of permutations of $Q$ with fermionic states, it is due to the fact that the BRS charge obeys anti-commutation relations with the fermionic creation operators. Consider now a generic physical state, containing one or more longitudinal particles as defined in the Equivalent Gauge (\ref{egL}). By applying eq.~(\ref{Qtens}), it can be written in terms of the ``standard'' longitudinal of eq.~(\ref{stdL}), plus one BRS-exact state
\beq
\displaystyle
|{\textrm{ph}}_{1}\rangle\ldots|W_L\rangle\ldots |{\textrm{ph}}_{n}\rangle
=|{\textrm{ph}}_{1}\rangle\ldots|w_0\rangle\ldots |{\textrm{ph}}_{n}\rangle
+\frac1{m}\, Q\left\{
|{\textrm{ph}}_{1}\rangle\ldots|\ob\rangle\ldots |{\textrm{ph}}_{n}\rangle
\right\}\,,
\eeq
where we used the fact that all the physical single particle states, with both the standard and the modified definition, are annihilated by the BRS charge, $Q|{\textrm{ph}}_{k}\rangle=0$. Moreover, since these states are all bosonic, $p_k=1$. By iteratively applying the above equation, all the $|W_L\rangle$'s can be converted into standard longitudinals up to exact states, and the equivalence is proved.

\subsection{Interacting Theory}\label{it}

The situation is not much different in the interacting theory, because there are several structural features of the free theory which are unaffected by the presence of interactions, as long as the latter remain perturbative. In the first place, perturbative interactions do not modify the particle content of the theory and the transformation properties of the particles under the Lorentz and the ghost number symmetry groups. Therefore like in the free case the single-particle states of the interacting theory are given by one spin one triplet $|w_h\rangle$ plus two scalars $|s\rangle$ and $|g\rangle$ with zero ghost number, and by the ghost/anti-ghost states $|\o\rangle$ and $|\ob\rangle$. Of course we have three identical replicas of such particles, forming triplets under the custodial group, we will omit the custodial index for shortness. Furthermore, the spectrum contains a physical Higgs scalar, which however will not play any role in what follows. 

The transformation properties of the states under the CPT symmetry is also unaffected by the interactions, therefore up to a phase choice the action of the CPT operator on our states is still given by eq.~(\ref{cptp}). The reader might be confused by this equation in the case of the interacting theory. The CPT operator should connect \emph{in} and \emph{out} asymptotic states, therefore it makes little sense to declare, as in eq.~(\ref{cptp}), that some states are CPT eigenvalues. However this is perfectly acceptable for single-particle states, because the \emph{in} and the \emph{out} states of a single isolated particle are just identical up to a relative phase choice. 

Naively, one would guess that another feature of the theory that interactions can not change is the norm of the Fock space, therefore the scalar products among single-particle states should still given by eq.~(\ref{nor}), with ${\mathcal{N}}$ as in Table~\ref{table:norm}, at most up to a linear redefinition of the states. Actually, this is too naive, eq.~(\ref{nor}) does not hold in general, but only when one particular condition is enforced on the gauge-fixing parameters $\mt$ and $\xi$. Already in the free case, indeed, eq.~(\ref{nor}) was derived under the assumption $\mt=m$ and $\xi=1$, and it would be violated for $\mt\neq m$. When $\mt\neq m$, as previously discussed, dipole terms appear in the propagators and the canonical particle interpretation no longer holds. This problem was raised long ago by K\"{a}ll\'en \cite{Kallen:1952zz} in the case of massless QED, where dipoles are present in non-Feynman gauges $\xi\neq1$, and solved many years later \cite{Nakanishi:1972sm,Becchi:1974md}. It turns out that a particle interpretation is possible even in the presence of dipoles, but it requires a modified LSZ formalism. In particular, in the approach of Ref.~\cite{Nakanishi:1972sm} the single-particle scalar products are modified, and contain more singular distributions besides the standard $\delta^3(\p-\q)$. Fortunately we will not need to deal with these complications because the dipole terms can be canceled, at all orders in perturbation theory, by a suitable condition on the gauge-fixing parameters, which we will derive in Section~\ref{mf}. Under this condition, eq.~(\ref{nor}) holds without subtleties.

Similar considerations apply to the condition of a completely degenerate mass spectrum, which was enforced in the free theory by choosing $\xi=1$. Notice that mass degeneracy is an essential ingredient for the formulation of the Equivalent Gauge already in the free case. If the mass of the scalar $|s\rangle$ was different from the one of the zero-helicity vector boson $|w_0\rangle$, the configuration-space wave function of the two states would have different frequencies and the shift (\ref{egL}) of the longitudinal state would not result in a simple shift of the wave function in momentum space as in eq.~(\ref{freg}). Thus, in this situation it would have been impossible to engineer the cancellation of the anomalous high energy behavior. In the interacting theory the pole mass of the particles will be renormalized by the radiative corrections, and there is no reason why the mass of the spin one states $|w_h\rangle$, {\it{i.e.}} the mass ``$m$'' of the physical $W$ particle, should renormalize in the same way as the mass of the scalars $|s\rangle$, $|g\rangle$, $|\o\rangle$ and $|\ob\rangle$. Therefore the degeneracy of the spectrum will not be preserved automatically by radiative corrections after being imposed at the tree-level. Instead, as we will show in the next section, what radiative corrections can not break is the degeneracy of the scalar particles among themselves, this property is ensured by the BRS symmetry. Since the common mass of the scalars is gauge-dependent, it is possible to set it equal to $m$ at each order in perturbation theory by choosing one combination of the two gauge-fixing parameters $\xi$ and $\widetilde{m}$.  This second condition, together with the one of dipole cancellation previously discussed, will determine the gauge-fixing parameters completely, in a way that differs from eq.~(\ref{gch}) beyond tree-level order.

The BRS charge $Q$ is also modified by radiative corrections, however its form is strongly constrained by the symmetries enumerated in Section~\ref{s11}. Let us start discussing how $Q$ acts on the single-particle states. Since $Q$ commutes with the $P_\mu$ operators, and no multi-particle state has the same $4$-momentum of a single-particle one, the action of $Q$ must be closed within the single-particle subspace. In practice, since $Q$ is a linear operator, this means that $Q$ acting on a single particle must be a linear combination of single-particle states. Moreover, the action of $Q$ is constrained by the fact that it is a Lorentz scalar and that it has ghost number equal to $+1$. By combining these informations one can show that it must have the generic form 
\bea
& Q|w_h\rangle\,=\,Q|\o\rangle\,=\,0\,,\;\;\;\;\; & Q|s\rangle\,=\,i\,q_{s\o}\,|\o\rangle\,,\nn\\
& Q|g\rangle\,=\,i\,q_{g\o}\,|\o\rangle\,, \;\;\;\;\;\;\;\;& Q|\ob\rangle\,=\,i\,q_{\ob s}\,|s\rangle
+\,i\,q_{\ob g}|g\rangle\,.
\label{brs0}
\eea
By further imposing the CPT symmetry we find that the four parameter appearing in the above equation, which a priori could have been complex, are actually purely real because $Q$ is a CPT-odd operator, as in eq.~(\ref{cptq}), while the scalar particle states are CPT-even as in eq.~(\ref{cptp}). Finally, we must take into account that $Q$ is Hermitian (\ref{her}) and nilpotent (\ref{nil}). These properties imply, respectively
\beq
\renewcommand{\arraystretch}{1}
\bigg\{\hspace{-2pt}\begin{tabular}{l}
$q_{\ob s}=-q_{s\o}$\\
$q_{\ob g}=q_{g\o}$
\end{tabular}
,\;\;\;\;\;\;\;\;\;\;\;\;\;
q_{\ob s} q_{s \o}+q_{\ob g} q_{g \o}=0\,.
\label{bc1}
\eeq
The first two conditions arise from imposing Hermiticity of the $Q$ matrix elements among single-particle states, notice the crucial sign difference among the two equations, which is due to the opposite norm of the scalar and of the Goldstone.

In summary, we have seen that the BRS operator restricted to the single-particle subspace can be parametrized by four real constants, and we have derived three relations among them. This allows us to determine $Q$ up to a multiplicative renormalization factor and also, since one of the equations is quadratic, up to a twofold sign ambiguity. However the sign ambiguity is immediately resolved by noticing that perturbative corrections can not cause a sign flip, therefore the signs in the BRS charge should match those of the free case in eq.~(\ref{brsfree}).\footnote{
Actually, the overall sign of the charge will not matter for us, we only have to determine the relative sign among $q_{\ob s}$ and $q_{\ob g}$, {\it{i.e.}} to decide, given eq.~(\ref{bc1}), whether $q_{\ob s}/q_{\ob g}=-q_{s\o}/q_{g\o}$ is equal to $+1$ or to $-1$.
}
 The final result can be written as
\bea
& Q_R|w_h\rangle\,=\,Q_R|\o\rangle\,=\,0\,,\;\;\;\;\; & Q_R|s\rangle\,=\,i\,m\,|\o\rangle\,,\nn\\
& Q_R|g\rangle\,=\,-i\,m\,|\o\rangle\,, \;\;\;\;\;\;\;\;& Q_R|\ob\rangle\,=\,-i\,m\,\left(|s\rangle+|g\rangle\right)\,,
\label{brsint}
\eea
where the renormalized charge is defined by
\beq
Q\,=\,Z_Q\,Q_R\,.
\eeq
Up to the renormalization constant $Z_Q$, which could be computed order by order in perturbation theory, we have found that the BRS charge of the interacting theory acts exactly like the free one of eq.~(\ref{brsfree}).

Until now we have restricted our attention to the states describing a single isolated particle, and it might seem very difficult to extend our analysis to multi-particle \emph{in} and \emph{out} asymptotic states. This is indeed a complicated problem, which has however a simple solution thanks to an important theorem proved in full generality by Kugo and Ojima \cite{KO}. The theorem states that any internal symmetry generator acts on the asymptotic states as the tensor product of the associated single particle representations. In the case of the BRS charge operator, this means \footnote{In the language of asymptotic fields adopted in \cite{KO}, the theorem is stated by saying that the symmetry generator is a quadratic polynomial in the asymptotic fields. This is completely equivalent to eq.~(\ref{KOth}).}
\beq
Q_R\,|\Psi_{1}\ldots\Psi_{n}\,{;\,}^{\scalebox{0.9}{\textrm{\emph{\,in}}}}_{\scalebox{0.9}{\textrm{\emph{out}}}}\rangle\,=\,\sum_{k=1,n}(-)^{p_k}
|\Psi_{1}\ldots\left[Q_R\,|\Psi_{k}\rangle\right]\ldots \Psi_{k}\,{;\,}^{\scalebox{0.9}{\textrm{\emph{\,in}}}}_{\scalebox{0.9}{\textrm{\emph{out}}}}\rangle\,.
\label{KOth}
\eeq
The same property was derived in the previous section, eq.~(\ref{Qtens}), for the free BRS charge. Notice that the result above is quite simple to prove when the conserved charge acts linearly on the field operators, but it is rather non-trivial when, as it happens for the BRS operator in Table~\ref{table:brs}, non-linear terms appear in the commutators of the charge with the fields.

We have found, in conclusion, that the renormalized BRS charge acts exactly like the free-theory one both on the single-particle (\ref{brsint}) and on the multi-particle states, thanks to the Kugo-Ojima theorem (\ref{KOth}). Therefore the classification of the physical states, defined as the cohomology of the $Q_R$ operator, can be carried on exactly like in the free case. As in the free case, then, the physical states could be represented in the standard way, through the \mbox{$|w_h\rangle$'s}, but one can as well employ alternative representatives by performing a shift with some BRS-exact state. Similarly to eq.~(\ref{egL}), we define the longitudinal state as
\beq
\displaystyle
|W_L\rangle=|w_0\rangle+\frac\lambda{m}\,Q_R\,|\ob\rangle=|w_0\rangle-i\lambda\big[|s\rangle+|g\rangle\big]\,,
\label{egLint}
\eeq
for an appropriate choice of the parameter $\lambda$, to be specified below.

The discussion presented up to now resembles very closely the one of the free theory, showing that indeed, as anticipated, the formulation of the Equivalent Gauge is not much different in the interacting case. To go on, again following the steps performed for the free theory, we must compute the Feynman rule associated with external longitudinal $W$'s, and show that the anomalous growth of the polarization vector can be canceled by a suitable choice of $\lambda$. Differently from the free one, in the interacting case we will find $\lambda\neq1$. The matrix elements with external \emph{in} or \emph{out} particles, among which the $S$-matrix elements, can be expressed in terms of Feynman amplitudes by the standard LSZ formalism.\footnote{The reduction formulas are a well-known fundamental result of Quantum Field Theory, which however are typically derived for a positive-norm Fock space of particles. For completeness, we briefly discuss in Appendix~\ref{RF} how the derivation changes in the presence of negative norm states.} Let us consider, for definiteness, the matrix element of some time-ordered product of local operators among the vacuum and a single incoming particle state, it is given by \footnote{We omit the \emph{in} or \emph{out} labels on the asymptotic single-particle states, remembering that the two are identical as previously discussed.}
\beq
\langle0|T\left\{\O_1\ldots\O_n\right\}|\Psi_I(\p)\rangle=
\sum_{\ih}\langle \O_1\ldots\O_n[\Phi^\dagger_\ih(p)]_A\rangle{\mathcal{M}}_{\ih I}\,,
\label{redform}
\eeq
where $\Phi_\ih$ denotes any of the fundamental fields of our theory, with the exception of the non-propagating auxiliary field $B$.\footnote{If not specified otherwise, in this paper we will always be dealing with bare fields and parameters, we omit the bare quantities label ``$\ _{0}$'' and reintroduce it only when needed. } The correlator on the right hand side of the above equation has been amputated with respect to the $\Phi^\dagger_\ih$ external leg and corresponds, in perturbation theory, to a sum of amputated Feynman diagrams. The external wave-function factors, encoded in the matrix $\mc{M}$, are given by matrix elements of the fields among the vacuum and the single-particle states
\beq
{\mathcal{M}}_{\ih I}\,\equiv\,\langle0|\Phi_\ih(0) |\Psi_I(\p)\rangle\,.
\label{1me}
\eeq
The result is the same as in the free case, even if a bit more of work was needed to derive it. The Feynman rule for external states is still provided by the single-particle amplitudes of the corresponding fields.

\renewcommand{\arraystretch}{1.5}
\setlength{\tabcolsep}{3.5pt}
\begin{table}[t]
\begin{center}
\begin{tabular}{rccccl}
\ & $|w_h\rangle$ &  $|s\rangle$ &  $|g\rangle$  \\
\multirow{2}{*}{$\mathcal{M}\,=\left( \vphantom{\begin{array}{c}
 +1 \\ +1 
\end{array}} \right.\hspace{-4pt}$} 
& $\displaystyle\rho_w^\bot\epsilon_\mu^h\;\;$
& $\displaystyle-\rho_w\epsilon_\mu^s\;$ & $\displaystyle\rho_w\gamma_{wg}\epsilon_\mu^s$ 
& \multirow{2}{*}{$
\hspace{-4pt}\left. \vphantom{\begin{array}{c}
 +1 \\ +1 
\end{array}} \right)\hspace{-4pt}
\begin{array}{c}
W_{\mu}\\  \pi
\end{array}$}\\
& $0\;\;$ & $\displaystyle-\rho_\pi\gamma_{\pi s}\;$ & $\displaystyle\rho_\pi$ & \\
\end{tabular}
\end{center}
\caption{The single-particle matrix elements of the bare fields. The polarization vectors $\epsilon_\mu^h$ and $\epsilon_\mu^s$ are defined in eq.~(\ref{pvec}).}
\label{table:prf}
\end{table}

However the single-particle amplitudes are more complicated than the free ones, reported in eq.~(\ref{renf}). By exploiting Lorentz and CPT symmetry, and focusing on the zero ghost number sector, the matrix $\mc{M}$ can be parametrized as in  Table~\ref{table:prf}, in terms of five real parameters $\rho_w^\bot$, $\rho_w$, $\rho_\pi$, $\gamma_{wg}$ and $\gamma_{\pi s}$. In the free case, $\rho_w^\bot$, $\rho_w$ and $\rho_\pi$ are all equal to $1$ while in general, in the presence of interactions, they are different and furthermore they diverge because of multiplicative renormalization of the $W$ and of the $\pi$ fields. Moreover, the single-particle amplitude matrix is diagonal in the free case, while non-vanishing (but finite) $\gamma_{wg}$ and $\gamma_{\pi s}$ are generated by the interactions. Given the definition (\ref{egLint}) of the state, we immediately compute the matrix element of a longitudinally polarized $W$
\bea
\langle0|T\left\{\O_1\ldots\O_n\right\}|W_L(\p)\rangle\,=\,&&\hspace{-18pt}
\left[\rho_w^\bot\epsilon_\mu^0(\p)+i\,\lambda\,\rho_w(1-\gamma_{wg}) \epsilon_\mu^s(\p)\right]
{\mathcal{A}}[W_\mu(p)]\nn\\
&&\hspace{-18pt}\,-\,i\,\lambda\,\rho_\pi(1-\gamma_{\pi s})\,{\mathcal{A}}[\pi(p)]\,,
\label{resu0}
\eea
where ${\mathcal{A}}[W]$ and ${\mathcal{A}}[\pi]$ denote, respectively, the amputated amplitudes with an external $W$ or $\pi$ leg. The anomalous energy growth of the polarization vector is canceled by choosing
\beq
\displaystyle
\lambda\,=\,\frac{\rho_w^\bot}{\rho_w}\frac1{1-\gamma_{wg}}\,.
\label{pch}
\eeq
Indeed with this choice eq.~(\ref{resu0}) becomes 
\beq
\langle0|T\left\{\O_1\ldots\O_n\right\}|W_L(\p)\rangle\,=\,
\sqrt{Z_W}\epsilon_\mu^L(\p)\,{\mathcal{A}}[W_\mu(p)]\,+\,\sqrt{Z_\pi}\epsilon_\pi(\p)\,{\mathcal{A}}[\pi(p)]\,,
\label{resu}
\eeq
where the polarization vectors are defined as in eq.~(\ref{freg}), {\it{i.e.}}
\bea
&&\epsilon_\mu^L(\p)= \epsilon_\mu^0(\p)+i\,\epsilon_\mu^s(\p)=-\frac{m}{E_\p+|\p|}
\left\{1,\,\frac{\p}{|\p|}\right\}\,,\nn\\
&&\epsilon_\pi(\p)=-i\,,
\label{freg1}
\eea
and the wave function renormalization factors, $\sqrt{Z_W}$ and $\sqrt{Z_\pi}$, are given by
\beq
\sqrt{Z_W}=\rho_w^\bot\,,\;\;\;\;\;\sqrt{Z_\pi}=\sqrt{Z_W}\frac{\rho_\pi(1-\gamma_{\pi s})}{\rho_w(1-\gamma_{wg})}\,.
\label{wff}
\eeq

Once again, the final result is a simple generalization of the free one. The matrix element is the sum of two terms, associated respectively to amputated Feynman diagrams with an external gauge or Goldstone legs, multiplied by the polarization vectors $\epsilon_\mu^L$ and $\epsilon_\pi$ as depicted in the lower part of Figure~\ref{frule}. The difference with the free case is that now the two terms are weighted by the wave function renormalization factors, to be computed ---with a given regulator, since they diverge--- at each order in perturbation theory according to eq.~(\ref{wff}). Needless to say, the divergences of $Z_W$ and $Z_\pi$ are needed to cancel those of the Feynman amplitudes, leading to a finite matrix element up to the renormalization of the bare couplings and masses and of the local operators $\O$. The result is trivially extended to outgoing particles, and to generic \emph{in} and \emph{out} states. Each external longitudinal state corresponds to two sets of diagrams with either a gauge or a Goldstone leg, weighted by the appropriate wave-function factors. Of course, conjugate wave functions have to be used for outgoing particles.

It is interesting to discuss the connection of our result with the Equivalence Theorem. The modified longitudinal polarization vector decreases, rather than growing, in the high energy limit. Therefore in that limit the gauge contribution is suppressed and the result is dominated by the Goldstone one which, going back to eq.~(\ref{egLint}), originated from the presence of the Goldstone state in the definition of the longitudinal $W$. Therefore in the high energy regime the longitudinal $W$ can be equivalently represented by a Goldstone boson, as stated by the Equivalence Theorem. However this is not completely correct, $|W_L\rangle$ is not equivalent to $|s\rangle$, but to $\lambda |s\rangle$, and $\lambda$ is different from one in the interacting theory. In its strict form \cite{Chanowitz:1985hj}, which states that the longitudinal amplitudes are equal to the Goldstone ones up to a phase, the Equivalence theorem is thus violated, the multiplicative correction factor $\lambda$ must be taken into account. That the Equivalence Theorem holds up to a multiplicative correction, {\it{i.e.}} $\lambda\neq1$ in our language, was first noticed in Ref.~\cite{Bagger:1989fc}, and is sometimes regarded as a problem, in the literature. Indeed a considerable amount of work has been done \cite{bulkET} trying to engineer a gauge-fixing and a renormalization scheme where the corrections disappear.\footnote{The parameter $\lambda$ depends on the scheme one chooses to connect the $\rho$ and $\gamma$ constants to physical observables, this makes even less interesting the debate on its value. However it is worth noticing that, though scheme-dependent, $\lambda$ is finite (upon renormalization of the bare couplings and masses) because the divergences in $\rho_w^\bot$ and $\rho_w$ both come from the one of the $W$ field and therefore they cancel in the ratio.} However $\lambda\neq1$ is not an issue, neither at the theoretical nor at the practical level. Indeed the value of $\lambda$ is not needed for any practical purpose, computing the amplitudes in the Equivalent Gauge, from which the high-energy limit is immediately obtained, only requires the knowledge of the wave-function renormalization factors $Z_W$ and $Z_\pi$ appearing in eq.~(\ref{wff}). The parameter $\lambda$ only enters in the intermediate steps of the derivation and it needs not to be computed explicitly to apply the result.

What instead needs to be computed, in a given scheme, are $Z_W$ and $Z_\pi$, but this is not much different than in any other QFT. Normally in QFT each physical states is excited from the vacuum by a single fundamental field, and thus a single wave function renormalization parameter appears in the formula for the scattering amplitudes. Here we have two of them simply because the state can be excited from the vacuum by two different fields. Having to deal with two renormalization constants rather than one seems to require additional work, but actually it does not because $Z_W$ and $Z_\pi$ are related by a very compact formula, which is derived as follows.\footnote{An alternative derivation will be presented in Section~\ref{mf}.} We read in Table~\ref{table:brs} that the auxiliary field $B$ can be expressed as the anti-commutator of the BRS charge with the anti-ghost $\ob$. By the EOM of $B$ in eq.~(\ref{auxeom}) this implies
\beq
\frac{i}\xi\langle0|(\partial_\mu W^\mu+\mt\,\xi\,\pi)\cdot Q|\ob\rangle=\langle0|\{Q,\,\ob\}\cdot Q|\ob\rangle=0\,.
\eeq
The matrix element vanishes because $Q$ annihilates both the vacuum on the left and the BRS-exact state $Q|\ob\rangle$ on the right. Thus, remembering eq.~(\ref{brsint}), we have
\beq
\langle0|\partial_\mu W^\mu\cdot\big[|s\rangle+|g\rangle\big]
+\mt\,\xi\,\langle0| \pi\cdot\big[|s\rangle+|g\rangle\big]=0\,.
\eeq
which gives
\beq
\begin{array}{c}
m\,\rho_w(1-\gamma_{wg})=\mt\,\xi\,\rho_\pi(1-\gamma_{\pi s})\,, \\
 \resizebox{10pt}{!}{{$\Downarrow$}} \\
\displaystyle\sqrt{Z_\pi}=\sqrt{Z_W}\frac{m}{\mt\,\xi}\,.
\end{array}
\label{zrel}
\eeq
By this relation, the result of the present section can be finally summarized in a rather compact formula
\beq
\displaystyle
\langle0|T\left\{\O_1\ldots\O_n\right\}|W_L(\p)\rangle\,=\,
\sqrt{Z_W}\left[\epsilon_\mu^L(\p)\,{\mathcal{A}}[W_\mu(p)]\,+\,\frac{m}{\mt\,\xi}\epsilon_\pi(\p)\,{\mathcal{A}}[\pi(p)]\right]\,.
\label{finres}
\eeq

\section{Mass Degeneracy and Dipole Cancellation}\label{mf}

\begin{figure}
\begin{center}
\includegraphics[width=0.8\textwidth]{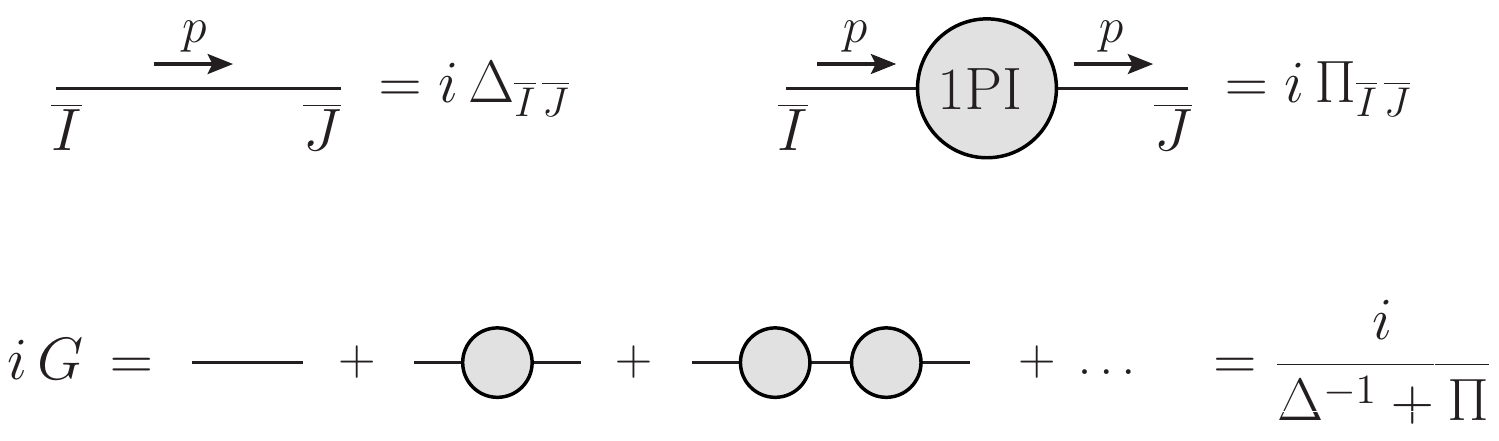}
\end{center}
\vspace{-15pt}
\caption{Graphical definition of the vacuum polarization amplitudes, of the Feynman propagators and of the two-point Green's functions.}
\label{gs}
\end{figure}

The analysis of the previous section relies on two conditions, dipole cancellation and mass degeneracy, which is now time to demonstrate. Actually, as already mentioned, we will see that these two conditions do not hold automatically but they need to be enforced by a suitable choice of the gauge-fixing parameters. This means that our main result, eq.~(\ref{finres}), only holds for a unique choice of $\mt$ and $\xi$, the one which ensures the validity of the conditions under which it is derived. Notice that this is not surprising because the left-hand side of the equation is gauge-independent, being a physical amplitude, while the right-hand one clearly is not. The two can thus be equal only for a given gauge choice.

\renewcommand{\arraystretch}{1.5}
\setlength{\tabcolsep}{3.5pt}
\begin{table}[t]
\begin{center}
\begin{tabular}{rcccl}
\ & $W_{\nu,\,b}$ &  $\pi_b$  \\
\multirow{2}{*}{$\Delta_B^{-1}\,=\delta_{ab}\times\left[ \vphantom{\begin{array}{c}
 +1 \\ +1 
\end{array}} \right.\hspace{-4pt}$} 
& $\displaystyle(m_0^2-p^2)\left(\eta_{\mu\nu}-\frac{p_\mu p_\nu}{p^2}\right)
+\left(m_0^2-\frac{p^2}\xi\right)\frac{p_\mu p_\nu}{p^2}$ & $i\,(m_0-\mt)p_\mu$ & \multirow{2}{*}{$
\hspace{-4pt}\left. \vphantom{\begin{array}{c}
 +1 \\ +1 
\end{array}} \right]\hspace{-4pt}
\begin{array}{c}
W_{\mu,\,a}\\  \pi_a
\end{array}$}\\
& $-i\,(m_0-\mt)p_\nu$ & $p^2-\mt^2\xi$ & \\
\end{tabular}
\begin{tabular}{rcccl}
\ & $\o_{b}$ &  $\ob_b$  \\
\multirow{2}{*}{$\Delta_F^{-1}\,=\delta_{ab}\times\left[ \vphantom{\begin{array}{c}
 +1 \\ +1 
\end{array}} \right.\hspace{-4pt}$} 
& $0$ & $\mt^2\xi-p^2$ & \multirow{2}{*}{$
\hspace{-4pt}\left. \vphantom{\begin{array}{c}
 +1 \\ +1 
\end{array}} \right]\hspace{-4pt}
\begin{array}{c}
\o_{a}\\  \ob_a
\end{array}$}\\
& $\mt^2\xi-p^2$ & $0$ & \\
\end{tabular}
\end{center}
\caption{The inverse Feynman propagators. We label as $\Delta_B$ and $\Delta_F$, respectively, the propagator matrices in the bosonic and fermionic sector.}
\label{table:delta}
\end{table}

The relevant objects to be studied, in order to discuss dipoles and masses, are the Green's functions in momentum space
\beq
i\,G_{\ih\,\jh}(p)=\displaystyle\int\hspace{-2pt}d^4x \,e^{ipx}\langle\Phi_\ih(x)\Phi^\dagger_\jh(0)\rangle\,,
\label{grfunc}
\eeq
where $\Phi=\{W^a,\pi^a,\o^a,\ob^a\}$ collectively denotes the bare fields the theory.\footnote{As usual, the physical Higgs field $h$ can be safely ignored because the mixed $h$-$W$ and $h$-$\pi$ correlators vanish thanks to custodial symmetry.} In perturbation theory, the Green's functions are computed in terms of the bare Feynman propagators, which we denote as $\Delta$, and of the \mbox{$1$PI} vacuum polarization amplitudes, $\Pi$, by summing the geometric series as sketched in Figure~\ref{gs}. The inverse Feynman propagator is immediately extracted from the free part of the Lagrangian (\ref{gfl}) and it is reported, for generic gauge-fixing parameter $\mt$ and $\xi$, in Table~\ref{table:delta}. In the table, $m_0=gv/2$ denotes the bare mass-term of the gauge fields, which in general does not coincide with the gauge-fixing mass parameter $\mt$. For $\mt\neq m_0$ the $\pi$-$W$ mixing does not cancel in the Lagrangian, leading to a non vanishing $\pi$-$W$ mixed propagator. The vacuum polarization matrix $\Pi$ could be computed in perturbation theory by evaluating the corresponding  \mbox{$1$PI} diagrams. By imposing Lorentz, custodial, Bose and ghost number symmetry it is simple to show that it can be parametrized in terms of few scalar form-factors, defined in Table~\ref{table:pita}. At all orders in perturbation theory the form-factor do not have pole singularities and moreover, since all the particles are stable in our theory, they will be purely real functions with no branch cuts. 

\renewcommand{\arraystretch}{1.5}
\setlength{\tabcolsep}{3.5pt}
\begin{table}[t]
\begin{center}
\begin{tabular}{rcccl}
\ & $W_{\nu,\,b}$ &  $\pi_b$  \\
\multirow{2}{*}{$\Pi_B\,=\delta_{ab}\times\left[ \vphantom{\begin{array}{c}
 +1 \\ +1 
\end{array}} \right.\hspace{-4pt}$} 
& $\displaystyle\Pi_{ww}^\bot(p^2)\left(\eta_{\mu\nu}-\frac{p_\mu p_\nu}{p^2}\right)
+\Pi_{ww}^L(p^2)\frac{p_\mu p_\nu}{p^2}$ & $i\,\Pi_{w\pi}(p^2)p_\mu$ & \multirow{2}{*}{$
\hspace{-4pt}\left. \vphantom{\begin{array}{c}
 +1 \\ +1 
\end{array}} \right]\hspace{-4pt}
\begin{array}{c}
W_{\mu,\,a}\\  \pi_a
\end{array}$}\\
& $-i\,\Pi_{w\pi}(p^2)p_\nu$ & $\Pi_{\pi\pi}(p^2)$ & \\
\end{tabular}
\begin{tabular}{rcccl}
\ & $\o_{b}$ &  $\ob_b$  \\
\multirow{2}{*}{$\Pi_F\,=\delta_{ab}\times\left[ \vphantom{\begin{array}{c}
 +1 \\ +1 
\end{array}} \right.\hspace{-4pt}$} 
& $0$ & $\Pi_{\o\ob}(p^2)$ & \multirow{2}{*}{$
\hspace{-4pt}\left. \vphantom{\begin{array}{c}
 +1 \\ +1 
\end{array}} \right]\hspace{-4pt}
\begin{array}{c}
\o_{a}\\  \ob_a
\end{array}$}\\
& $\Pi_{\o\ob}(p^2)$ & $0$ & \\
\end{tabular}
\end{center}
\caption{The \mbox{$1$PI} vacuum polarization amplitudes.}
\label{table:pita}
\end{table}

In order to obtain the Green's functions we just have to sum up $\Delta^{-1}$ and $\Pi$ and compute the inverse, as in Figure~\ref{gs}. The matrix of the bosonic sector is not completely trivial to invert, to simplify the calculation we rewrite it as
\beq
G_B^{-1}=\Delta^{-1}_B+\Pi_B=
A(p^2)\,{\mc{P}}_\bot+{\mc{P}}^iX_i^{\;j}{\mc{P}}^\dagger_j\,,
\eeq
where ${\mc{P}}_\bot$ is the transverse projector and ${\mc{P}}^{1,2}$ are ``longitudinal'' $5$-vectors
\beq
\displaystyle
{\mc{P}}_\bot=\left(
\begin{array}{cc}
\displaystyle \eta_{\mu\nu}-\frac{p_\mu p_\nu}{p^2} & 0 \\
0 & 0
\end{array}
\right)
\,,\;\;\;\;\;{\mc{P}}^{1}=
\left(\begin{array}{c}
\displaystyle i\frac{p_\mu}{p}\\0
\end{array}\right)
\,,\;\;\;\;\;
{\mc{P}}^{2}=
\left(\begin{array}{c}
0\\1
\end{array}\right)
\,.
\eeq
The $2\times2$ form factor matrix $X$ is given by
\beq
X=\left(
\begin{array}{cc}
\displaystyle B(p^2)-\frac{p^2}\xi & p\left[C(p^2)-\mt\right]\\
p\left[C(p^2)-\mt\right] & p^2F(p^2)-\mt^2\xi
\end{array}
\right)\,,
\eeq
where for shortness, and with the aim of matching the notation of Ref.~\cite{KO}, we have defined the form-factors $A$, $B$, $C$ and $F$ as
\beq
\begin{array}{ll}
A(p^2)=m_0^2-p^2+\Pi_{ww}^\bot(p^2)\,,\;\;\;\;&B(p^2)=m_0^2+\Pi_{ww}^L(p^2)\,,\\
C(p^2)=m_0+\Pi_{w\pi}(p^2)\,,\;\;\;\;&p^2F(p^2)=p^2+\Pi_{\pi\pi}(p^2)\,.
\end{array}
\eeq
In this parametrization the two-point function matrix is simply expressed as
\beq
G_B=
A^{-1}{\mc{P}}_\bot+{\mc{P}}^i\left(X^{-1}\right)_i^{\;j}{\mc{P}}^\dagger_j\,,
\label{twop}
\eeq
in terms of the inverse of $X$.

By using symmetries we have significantly constrained the Green's function matrix, showing that it can be parametrized in terms of few form-factors. We can do more if we further impose BRS invariance, which leads to certain Slavnov-Taylor identities for the two-point functions, proved in appendix~\ref{ST}. The first identity (\ref{ST1}) is a linear constraint on the bosonic propagators
\bea
\displaystyle
\partial_\mu^x\partial_\nu^y
\langle W^{\mu,\,a}(x) W^{\nu,\,b}(y)
\rangle
+\mt\,\xi\,\partial_\mu^x
\langle W^{\mu,\,a}(x) \pi^{b}(y)
\rangle&&\nn\\
+\mt\,\xi\,\partial_\nu^y
\langle \pi^{a}(x) W^{\nu,\,b}(y)
\rangle
+\mt^2\,\xi^2\,\langle \pi^{a}(x) \pi^{b}(y)
\rangle&&\hspace{-15pt}=-\,i\,\xi\,\delta^{ab}\delta^4(x-y)\,.
\label{ST1tx}
\eea
After going to Fourier space and substituting eq.~(\ref{twop}) this gives 
\beq
\displaystyle
p^2(X^{-1})_{11}+2\mt\,\xi\,p(X^{-1})_{12}+\mt^2\xi^2(X^{-1})_{22}\,=\,-\,\xi\,,
\label{brsprp}
\eeq
which, by computing explicitly the inverse, results in a very simple relation among the form-factors
\beq
B(p^2)F(p^2)=C^2(p^2)\,,
\label{korel}
\eeq
in agreement with Ref.~\cite{KO}, where the same result was derived from the Slavnov-Taylor identities for the $1$PI effective action. 

The physical masses corresponds to poles in the Green's functions, let us discuss where these divergences occur. The first term in eq.~(\ref{twop}), proportional to the transverse projector, is associated with the propagation of the spin one states $|w_h\rangle$. It diverges when $A$ vanishes, and therefore the physical $W$ boson mass $m$ is defined implicitly by
\beq
A(m^2)\,=\,m_0^2-m^2+\Pi_{ww}^\bot(m^2)\,=\,0\,.
\label{wmass}
\eeq
In general, $A$ has a simple zero at $p^2=m^2$, leading to a simple pole in the propagator. The second term in eq.~(\ref{twop}) diverges when the matrix $X$ becomes singular, $\textrm{Det}[X]=0$, and its singularities correspond to the masses of the scalar and of the Goldstone states. A priori, the two masses could be different because the pole condition $\textrm{Det}[X]=0$ is quadratic in the form-factors and thus it could admit two distinct solutions in $p^2$. However, thanks to the BRS relation of eq.~(\ref{korel}), the determinant can be expressed as a perfect square 
\beq
\textrm{Det}\left[X\right]\,=\,-\,\frac1{\xi B}\left(\mt\,\xi\,B-p^2C\right)^2\,
\label{det}
\eeq
so that the pole condition is actually linear
\beq
\mt\,\xi\,B(p^2)=p^2C(p^2)\,.
\label{sgm}
\eeq
At each order in perturbation theory, the equation above has a unique solution and thus we can conclude that, because of BRS symmetry, the scalar and the Goldstones are degenerate. This is of course not surprising because they are part of the same multiplet of the conserved BRS charge as in eq.~(\ref{brsint}). The scalar and the Goldstones thus have a common mass, defined by eq.~(\ref{sgm}), which however is different, in general, from the one of the $W$'s given by eq.~(\ref{wmass}). But the scalar and Goldstone mass is gauge-dependent, thus we can set it equal to $m$ by a gauge condition, namely we impose that eq.~(\ref{sgm}) is satisfied at $p^2=m^2$, {\it{i.e.}}
\beq
\mt\,\xi\,B(m^2)=m^2C(m^2)\;\;\;\resizebox{16pt}{!}{{$\Leftrightarrow$}}\;\;\;\mt\,\xi\,=\,m^2\frac{m_0+\Pi_{w\pi}(m^2)}{m_0^2+\Pi^L_{ww}(m^2)}\,.
\label{gaugecond1}
\eeq
At the tree-level order, where $\Pi=0$ and $m_0=m$, the above condition reduces to the familiar $\mt\,\xi=m$, which indeed ensures the degeneracy of the particle spectrum. 

The second condition we need, besides mass-degeneracy, is the cancellation of dipoles. Also in this condition is not automatic, dipole terms are generically present in the propagator because the determinant of $X$ (\ref{det}) is a perfect square, and therefore it has a double zero at $p^2=m^2$. The $1/\textrm{Det}[X]$ factor in $X^{-1}$ will thus lead to double poles. Indeed after some manipulation, and making use of the relation
\beq
 \left(\mt\,\xi\,B-p^2C\right) = -\, \frac{B}{C} \left(p^2F-\mt\,\xi\,C\right)\,,
\label{id1}
\eeq
which easily follows from eq.~(\ref{korel}), $X^{-1}$ can be explicitly written as
\beq
X^{-1}=\left(
\begin{array}{cc}
\displaystyle 
\frac{-\,\xi\,F}{p^2F-\mt\,\xi\,C}\left[1+\frac{\mt\,\xi\,(C-\mt)}{p^2F-\mt\,\xi\,C}\right]
 &
 \displaystyle \frac{-\xi\,p\,C(C-\mt)}{(\mt\,\xi\,B-p^2C)(p^2F-\mt\,\xi\,C)}\\
\displaystyle
\frac{-\xi\,p\,C(C-\mt)}{(\mt\,\xi\,B-p^2C)(p^2F-\mt\,\xi\,C)} & 
\displaystyle 
\frac{-\,B/\mt}{\mt\,\xi\,B-p^2C}\left[1+\frac{p^2(C-\mt)}{\mt\,\xi\,B-p^2C}\right]
\end{array}
\right)\,.
\label{xinv}
\eeq
Because of eq.s~(\ref{gaugecond1}) and (\ref{id1}), each of the factors in the denominators have one simple zero around $p^2=m^2$, and therefore each entry of the matrix displays a double pole singularity. However we notice that these double poles are all proportional to $C-\mt$, so that we can cancel them by choosing $C=\mt$ at the pole. We thus obtain the second gauge-fixing condition
\beq
\mt\,=\,C(m^2) \;\;\;\resizebox{16pt}{!}{{$\Leftrightarrow$}}\;\;\;\mt=m_0+\Pi_{w\pi}(m^2)\,.
\label{gaugecond2}
\eeq
At tree-level, the equation above reduces to the usual $R_\xi$-gauge condition $\mt=m$, which indeed cancels the double poles in the free propagators. 

By combining eq.s~(\ref{gaugecond1}) and (\ref{gaugecond2}), the two gauge-fixing conditions which ensure dipole cancellation and mass degeneracy are finally rewritten as
\beq
\left\{
\begin{array}{l}
\displaystyle
\mt=m_0+\Pi_{w\pi}(m^2)\\
\displaystyle
\xi=\frac{m^2}{m_0^2+\Pi_{ww}^L(m^2)}
\end{array}
\right.\,.
\label{gffin}
\eeq
The two gauge-fixing parameters are now completely determined and can be computed, in a given renormalization scheme, at all orders in perturbation theory.\footnote{Obviously the form factors themselves depend on $\mt$ and $\xi$, so that eq.~(\ref{gffin}) is a system of implicit equations, to be solved perturbatively.}

With the gauge choice of eq.~(\ref{gffin}), and only with that one, the assumptions we made in Section~\ref{it} are verified, and thus the final result (\ref{freg1}) is correct. However in order to apply it to a practical calculation we still have do determine the two wave-function renormalization parameters $Z_W$ and $Z_\pi$, let us see how to extract them from the Green's functions starting from their definition in eq.~(\ref{wff}). Actually, it will be sufficient to compute ${Z_W}$, because $Z_\pi$ is simply related to it by eq.~(\ref{zrel}), however it is interesting to discuss both of them and to re-derive eq.~(\ref{zrel}) in the present formalism. To proceed, we notice that now that dipoles are canceled and the degeneracy of the spectrum is enforced, all the Green's functions have simple poles at $p^2=m^2$, whose residues are related with the single-particle particle matrix elements of the corresponding fields defined in eq.~(\ref{1me}). Namely, as reported in eq.~(\ref{gf}) of Appendix~\ref{RF}, we have
\beq
\displaystyle
G_{\ih\,\jh}(p)
\underset{p^2\rightarrow m^2}{\xrightarrow{\hspace*{17pt}}}
\frac{1}{p^2-m^2}\,{\mc{M}}_{\ih I}\mc{N}^{IJ}{\mathcal{M}}^\dagger_{ J \jh}
\,,
\eeq
with the norm matrix $\mc{N}$ of Table~\ref{table:norm}. Using the parametrization of Table~\ref{table:prf} for $\mc{M}$, and remembering the completeness relation (\ref{crel}) of the polarization vectors, the bosonic propagator at the pole becomes
\beq
\displaystyle
G_B(p)
\underset{p^2\rightarrow m^2}{\xrightarrow{\hspace*{17pt}}}
\frac{1}{p^2-m^2}\left[
-({\mc{\rho}^\bot_{w}})^2\,\mc{P}_\bot-\mc{P}^i\left(M^t\sigma_3 M\right)_{i}^{\; j}\mc{P}^\dagger_j
\right]\,,
\eeq
where the matrix $M$ is defined as
\beq
M=\left(
\begin{array}{cc}
\displaystyle \rho_w & \rho_\pi\gamma_{\pi s}\\
-\rho_w\gamma_{wg}  &  -\rho_\pi
\end{array}
\right)\,.
\eeq
From the above equation, by comparing with eq.~(\ref{twop}) expanded around the pole, we are now able to express the $\rho$ and $\gamma$ parameters, and eventually $Z_W$ and $Z_\pi$, in terms of the form factors, {\it{i.e.}}
\beq
\displaystyle
Z_W=(\rho_w^\bot)^2=-\left(\frac{dA}{dp^2}\right)_{{p^2=m^2}}^{-1}\,,\;\;\;\;\;M^t\sigma_3 M=-\lim_{p^2\to m^2}\left[(p^2-m^2)X^{-1}\right]\,.
\label{eeqr}
\eeq
Not surprisingly, the wave-function renormalization of the gauge field is given by the derivative of the transverse form-factor, {\it{i.e.}}
\beq
Z_W=\left(1-\frac{d\Pi_{ww}^\bot}{dp^2}\right)_{{p^2=m^2}}^{-1}\,.
\eeq

Computing $Z_\pi$ is a bit more complicated. The best way to proceed is to start from the BRS relation of eq.~(\ref{brsprp}) for the matrix $X^{-1}$ and, by going at the pole, to convert it in an equation for $M$ by eq.~(\ref{eeqr}). The result is simply
\beq
\displaystyle
\left[
\frac{m\,\rho_w}{\sqrt{\xi}}(1-\gamma_{wg}) - \mt\sqrt{\xi}\rho_\pi(1-\gamma_{\pi s}) \right]
\left[
\frac{m\,\rho_w}{\sqrt{\xi}}(1+\gamma_{wg}) + \mt\sqrt{\xi}\rho_\pi(1+\gamma_{\pi s})
\right] = 0\,.
\label{eee}
\eeq
In order to decide which one of the two factors has to vanish, we notice that at the tree-level order, where $\rho_w=\rho_\pi=1$, $\mt=m$, $\xi=1$ and the $\gamma$'s vanish, the first term is zero while the second one is equal to $2\,m$. But perturbative loop corrections can not cancel the tree-level term, thus we can safely assume that the second factor will be different from zero at all orders in perturbation theory, and what vanishes is the first one.\footnote{By the ordinary renormalization theorems of the massive gauge theory, we know that the combinations $\rho_W/\sqrt{\xi}$ and  $\mt\sqrt{\xi}\rho_\pi$ are finite quantities, therefore all the terms in eq.~(\ref{eee}) are finite and the perturbative loop corrections are truly small changes of their tree-level values.} This gives exactly eq.~(\ref{zrel})
 \beq
\sqrt{Z_\pi}=\sqrt{Z_W}\frac{\rho_\pi(1-\gamma_{\pi s})}{\rho_w(1-\gamma_{wg})}=\sqrt{Z_W}\frac{m}{\mt\,\xi}\,,
 \eeq
which was derived in Section~\ref{it} by operatorial identities.

With the above result, our illustration of the Equivalent Gauge is basically complete. We have derived the two gauge-fixing conditions that ensure dipole cancellation and mass degeneracy and we have obtained explicit formulas for the wave-function renormalization parameters which appear in the Feynman rule for the longitudinal $W$'s. By now, the Equivalent Gauge is fully  specified. However there is one last point to be addressed, which we left behind in the discussion of mass degeneracy. We have shown that the Goldstones are degenerate with the scalars, but also the ghosts and the anti-ghost should have the same mass, because the four states together should form a multiplet of the conserved BRS charge, as in eq.~(\ref{brsint}). In order to show that this is indeed the case we will obviously need to impose BRS invariance, we will do that by the second Slavnov-Taylor identity (\ref{ST2}) we derived in Appendix~\ref{ST}. First of all we notice that, independently of BRS, the ghosts and the anti-ghosts must have the same mass, $m_\o$, which is defined, using Tables~\ref{table:delta} and \ref{table:pita}, by the condition
\beq
\mt^2\xi-m_\o^2+\Pi_{\o\ob}(m_\o^2)=0\,.
\eeq
Second, we impose eq.~(\ref{ST2}), that reads
\beq
\displaystyle
\xi\langle \O^a(x)\ob^b(y)
\rangle=\partial_\mu^x\partial_\nu^y
\langle W^{\mu,\,a}(x) W^{\nu,\,b}(y)
\rangle
+\mt\,\xi\,\partial_\mu^x
\langle W^{\mu,\,a}(x) \pi^{b}(y)
\rangle\,,
\label{last}
\eeq
where $\O^a(x)$, whose explicit form can be read from eq.~(\ref{ST2}), is a certain local operator with unit ghost number. In Fourier space, by using the LSZ reduction formula in eq.~(\ref{LSZ0}), we see that the correlator on the left-hand side has a pole at $m_\o$, {\it{i.e.}}
\beq
\displaystyle\int\hspace{-2pt}d^4x \,e^{-ipx}\langle\O(0)\ob(x)\rangle\;\;\underset{p^2\rightarrow m^2}{\xrightarrow{\hspace*{20pt}}}\frac{-i}{p^2-m_\o^2}\langle0|\O(0)|\ob_I(\p)\rangle\langle\o(\p)|\ob(0)|0\rangle\,.
\eeq
But on the right hand side of eq.~(\ref{last}) we have the bosonic propagators, whose poles are at $p^2=m^2$. By matching the poles we find
\beq
m_\o=m\,,
\eeq
and our derivation is complete.

\section{Conclusions}\label{conc}

I have described a novel covariant formulation of massive gauge theories where, differently from the ordinary one, the longitudinal polarization vectors do not grow with the energy. This renders the energy and coupling power-counting of the scattering amplitudes completely transparent at the level of individual Feynman diagrams. The main result is reported in eq.~(\ref{finres}), which provides the Feynman rule associated, in our formalism, with longitudinally polarized external vector bosons. As pictorially represented in Fig.~\ref{frule}, the scattering amplitude is represented by two sets of Feynman diagrams, one with gauge and the other with Goldstone external fields.\footnote{A similar notation has been proposed also in Ref.~\cite{Veltman:1989ud}.} The gauge amputated amplitude is weighted by a modified longitudinal polarization vector $\epsilon_\mu^L$, see eq.~(\ref{freg1}), which vanishes at high energy rather than growing like the standard one. Our result is particularly simple at the tree-level order, where the wave-function renormalization factor $Z_W$ is equal to one and the gauge-fixing parameters, reported in eq.~(\ref{gffin}) are $\mt=m_0=m$ and $\xi=1$. 

Since the energy power-counting is transparent, proving the Equivalence theorem ---which controls the high-energy behavior of the amplitudes--- must be straightforward in our formalism. Indeed $\epsilon_\mu^L\sim m/E$, while $\epsilon_\pi=-i$, thus in eq.~(\ref{freg1}) the Goldstone diagrams dominate in the high energy limit and the Equivalence Theorem is immediately demonstrated.\footnote{What we prove is actually a ``weak'', but equally powerful, form of the Equivalence Theorem, because the longitudinal matrix element is only proportional to the Goldstone amplitude, with a calculable proportionality factor $\frac{m}{\mt\xi}$ which differs from one beyond tree-level. This of course was to be expected by previous results on the Equivalence Theorem \cite{Bagger:1989fc}, as we discussed in detail in Section~\ref{it}.} Since our formalism makes the Equivalence Theorem is self-evident, we call it an ``Equivalent Gauge''.

One way to summarize the present paper is that it provides a simpler and arguably more physical proof of the Equivalence Theorem. However there is something more than that, because the Equivalent Gauge is an exact reformulation of the theory and thus it allows one to estimate, or even to compute if needed, the finite energy corrections in a systematic way. Indeed while it is definitely true that the gauge term in the amplitude is systematically suppressed by $m/E$ because of the polarization vector, its contribution relative to the Goldstone one needs not to be of order $m/E$. Extra sizable enhancement or suppression factors can arise because the gauge and the Goldstone diagrams contain, in general, completely different Feynman vertices which could describe interactions of completely different strength. For example it might happen that the gauge amplitude receives contributions from some large coupling, which is instead not present in the Goldstone one. In this case the relative importance of the gauge will be enhanced by the corresponding ratio of couplings. It might also happen, already in the SM as discussed in Footnote~\ref{foot}, that the Goldstone amplitude is suppressed by some additional $m/E$ power, compared to the gauge one, because of selection rules. In this case the $m/E$ suppression from the polarization vectors is compensated and the Equivalence Theorem is violated, while the Equivalent Gauge can straightforwardly deal with this situation.

As discussed in the Introduction, the Equivalent Gauge might be useful in all the problems where an energy or coupling power-counting needs to be set up. This for sure includes the formal proof of the Effective $W$ Approximation, but perhaps also other issues related with soft/collinear EW Sudakov resummations \cite{SEW}. For what concerns practical calculations, the obvious advantage of the Equivalent Gauge is that it allows to select, through power-counting, the most relevant Feynman diagrams for a given process, simplifying the calculation. Furthermore, differently from ordinary covariant gauges, the high-energy limit can be taken on each individual Feynman diagram, neglecting the mass terms in the propagators. For practical purposes, it is important to stress that in the Equivalent Gauge the Feynman rules for vertices and for internal line propagators are simply those of the ordinary $R_\xi$ gauges, what changes is just the Feynman rule for external longitudinal particles. Therefore all the existing calculation tools can be applied to the Equivalent Gauge with extremely mild modifications.

For definiteness, and with the aim of keeping the discussion as simple as possible, we have illustrated the Equivalent Gauge in the simplest weakly-coupled ({\it{i.e.}}, renormalizable) massive gauge theory, the \mbox{SU$(2)$} Higgs-Kibble model. However it is rather clear that our derivation does not rely on these details, it could be straightforwardly generalized to models with more particles and interactions. Instead, it would not be completely straightforward to generalize our results to the SM, because in the case of the Higgs-Kibble model the proof takes advantage of the presence of the unbroken custodial group \mbox{SO$(3)_c$}, which is broken in the SM by the gauging of Hypercharge. By custodial symmetry we could treat the gauge and the Goldstone fields and the states as identical replicas, and this has been an import an simplification in our analysis. Generalizing the results to the SM is definitely possible, but is left for future work because it requires to deal with some additional technical complication.

\subsubsection*{Note Added}

After this work was completed, Prof.~R.~Ferrari drew my attention on his paper, Ref.~\cite{Ferrari:2011bx}, where the possibility of shifting the longitudinal by a BRS-exact state, as in eq.~(\ref{egLint}), was mentioned as a way to prove the Equivalence Theorem. However the structure of the asymptotic states, the action of the interacting BRS charge on them and the way in which their matrix elements are related to Feynman diagrams by LSZ reduction formulas were not discussed in Ref.~\cite{Ferrari:2011bx}, while we have seen that these aspects are crucial, and highly non-trivial, in order to demonstrate that the shift (\ref{egLint}) leads to the modified Feynman rule in eq.~(\ref{finres}). Furthermore, in order to establish the result it is essential that the dipole cancellation condition is enforced on the gauge-fixing parameters at all orders in perturbation theory, a possibility which has been overlooked in Ref.~\cite{Ferrari:2011bx}. Also the second gauge-fixing condition, of degenerate spectrum, is not mentioned in Ref.~\cite{Ferrari:2011bx}, while the latter condition is crucial in order to obtain well-behaved polarization vectors. As previously explained, if the different states in eq.~(\ref{egLint}) had different masses the redefinition would not result in a shift of the amplitude in Fourier space, as the two states oscillate at different frequencies. Stated in a different way, the scattering amplitudes would correspond to poles of the correlators at different locations, to be computed by distinct sets of Feynman diagrams with different kinematics. The anomalous energy behavior of the polarization vectors would show up in each of the two sets and there would be no way to cancel it.

\section*{Acknowledgments}
I thank R.~Rattazzi and P.~Lodone for collaboration at the initial stage of this project, and C.~M.~Becchi for reading the manuscript. I also thank G.~Panico for reading a preliminary version of the manuscript and for helping me in improving the presentation. Finally, I acknowledge R.~Torre and P.~A.~Grassi for discussions. This work is partially supported by the  {\sc{MIUR-FIRB}} grant {\sc{RBFR12H1MW}}, by the grant SNF Sinergia {\sc{CRSII2-141847}} and by the {\sc{ERC}} Advanced Grant No. 267985 \emph{DaMeSyFla}.

\appendix

\section{The Reduction Formula}\label{RF}

The ordinary LSZ reduction formula relates the matrix elements on asymptotic states to the residues of the momentum-space field correlators at the pole masses. Of course it is a very standard QFT result, which however is usually derived in the case of a positive-norm Fock space of particles. The aim of this Appendix is to discuss how it gets modified for a generic norm matrix, such as the one we encountered in Table~\ref{table:norm}. The standard proof of the LSZ formula makes use of the completeness relation, and in particular of the completeness relation restricted to the single-particle subspace. But if the single particle states have a norm as in eq.~(\ref{nor}), the standard completeness relation gets modified and becomes
\beq
\left(\I\right)_{1\textrm{--particle}}=\sum_{I,J}\int\hspace{-2pt}\frac{d^3p}{2E_p}\mc{N}^{IJ}|\Psi_I(\p)\rangle\langle\Psi_J(\p)|\,,
\label{comprel}
\eeq
where $\mc{N}^{IJ}$ denotes the inverse of the norm matrix $\mc{N}_{IJ}$, which just coincides with $\mc{N}$ in the case of Table~\ref{table:norm}. In the above equation it is assumed that all the particles have the same mass $m$, so that $E_\p=\sqrt{\p^2+m^2}$ for all particle species.

To proceed, let us just repeat the steps through which the reduction formula is proven in the standard case. One considers the Fourier transform of a generic time-ordered correlator with respect to one of its variables, {\it{i.e.}}
\beq
\int\hspace{-2pt}d^4x\, e^{-iqx}\langle0|T\left\{\O_1(y_1)\ldots\O_n(y_n)\Phi^\dagger_\ih(x)\right\}|0\rangle\,,
\eeq
where we assume $\Phi_\ih(x)$ to be one of the fundamental fields of our theory, while the $\O$'s can be taken to be generic local operators, either fundamental or composite. One can prove that the only poles of the momentum-space correlator can arise from the integration region in configuration space where $x$ is in the far past, $x^0>y_i^0\;\forall i$, or in the far future, $x^0>y_i^0\;\forall i$. The two configurations will correspond to matrix elements involving incoming and outgoing particles, respectively. Let us consider incoming particles for definiteness. We have to focus on the far past region, where $\Phi_\ih(x)$ remains at the extreme right of the operator string. By inserting the completeness relation, which is now given by eq.~(\ref{comprel}), and performing the $x$ integral, we can derive the structure of the propagator near the pole. The incoming particle singularity arises when the external momentum $q$ becomes equal to the on-shell momentum of some physical particles, $q=p$, $p^2=m^2$, where the correlator takes the form
\bea
&\displaystyle\int\hspace{-2pt}d^4x \,e^{-iqx}\langle0|T\left\{\O_1\ldots\O_n\Phi^\dagger_\ih(x)\right\}|0\rangle\;\;\underset{q\rightarrow p}{\xrightarrow{\hspace*{30pt}}}&\nn\\
&\displaystyle\frac{i}{q^2-m^2}\langle0|T\left\{\O_1\ldots\O_n\right\}|\Psi_I(\p)\rangle
\mc{N}^{IJ}\langle\Psi_J(\p)|\Phi^\dagger_\ih(0)|0\rangle\,.&
\label{LSZ0}
\eea
We see that the matrix element of the time-ordered product $T\left\{\O_1\ldots\O_n\right\}$ among the vacuum and a single incoming particle can be computed as the residue at the pole of the $\O_1\ldots\O_n\Phi_\ih$ vacuum correlator. A similar expression holds for outgoing particles matrix elements
\bea
&\displaystyle\int\hspace{-2pt}d^4x \,e^{iqx}\langle0|T\left\{\Phi_\ih(x)\O_1\ldots\O_n\right\}|0\rangle\;\;\underset{q\rightarrow p}{\xrightarrow{\hspace*{30pt}}}&\nn\\
&\displaystyle\frac{i}{q^2-m^2}\langle0|\Phi_\ih(0)|\Psi_I(\p)\rangle \mc{N}^{IJ}
\langle \Psi_I(\p)|T\left\{\O_1\ldots\O_n\right\}|0\rangle 
\,.&
\label{LSZ1}
\eea

Generalizing the results above to arbitrary \emph{in} and \emph{out} multiparticle states is not completely straightforward, however the difficulties are just the same one encounters in the standard case of a positive-definite norm, and can be solved in the same way. The result is a very simple generalization of eq.s~(\ref{LSZ0}), (\ref{LSZ1}). Any \emph{in} or \emph{out} external particle will lead, respectively, to one more $\Phi$ and $\Phi^\dagger$ field insertion. The matrix elements on external multi-particle states, multiplied by the same factors as in eq.s~(\ref{LSZ0}), (\ref{LSZ1}), will provide the residual of the multiple pole singularity encountered when all the momenta go on-shell. The only difference with respect to the ordinary LSZ formula is in the norm matrix factor $\mc{N}$ which multiplies the single-particle matrix elements $\langle\Psi|\Phi^\dagger|0\rangle$ and $\langle0|\Phi|\Psi\rangle$, in the ordinary case $\mc{N}=\I$.

As an application of eq.s~(\ref{LSZ0}), (\ref{LSZ1}), consider the momentum-space Green's function of two fundamental fields
\beq
i\,G_{\ih\,\jh}(p)=\displaystyle\int\hspace{-2pt}d^4x \,e^{ipx}\langle0|T\left\{\Phi_\ih(x)\Phi^\dagger_\jh(0)\right\}|0\rangle\,.
\label{gf0}
\eeq
It has a pole at $p^2\rightarrow m^2$, of the form
\beq
\displaystyle
G_{\ih\,\jh}(p)
\underset{p^2\rightarrow m^2}{\xrightarrow{\hspace*{17pt}}}
\frac{1}{p^2-m^2}\,\langle0|\Phi_\ih(0)|\Psi_I(\p)\rangle\,\mc{N}^{IJ}\langle \Psi_J(\p)|\Phi^\dagger_\jh(0)|0\rangle\,.
\label{gf}
\eeq
This of course is just the usual statement that the particle mass is the location of the Green's functions poles and that the residues at the pole give the matrix elements of the fields among the vacuum and the single particle states. Since these single-particle amplitudes appear very often it is convenient to give them a name, we define
\beq
{\mathcal{M}}_{\ih I}\,\equiv\,\langle0|\Phi_\ih(0) |\Psi_I(\p)\rangle\,,\;\;\;\;\;
{\mathcal{M}}^\dagger_{ I \ih}\,\equiv\,\langle \Psi_I(\p) |\Phi_\ih^\dagger(0) |0\rangle\,.
\eeq

The calculation of the matrix elements can be greatly simplified if we introduce amputated field correlators. Any correlator with fundamental fields on the external legs takes, in momentum space, the factorized form
\bea
&&\int\hspace{-2pt}d^4x\, e^{-iqx}\langle0|T\left\{\O_1\ldots\O_n\Phi^\dagger_\ih(x)\right\}|0\rangle\,=\,\sum_\jh\, \langle \O_1\ldots\O_n[\Phi^\dagger_\jh(p)]_A\rangle\,i\,G_{\jh\,\ih}(p)\,,\nn\\
&&\int\hspace{-2pt}d^4x\, e^{-iqx}\langle0|T\left\{\Phi_\ih(x)
\O_1\ldots\O_n\right\}|0\rangle\,=\,\sum_\jh\, 
i\,G_{\ih\,\jh}(p)\,
\langle [\Phi_\jh(p)]_A\O_1\ldots\O_n\rangle\,,
\eea
where the correlators have been amputated by removing the propagators, including radiative corrections, on the external leg. The latter are collected in the Green's function factors. In the above equations, the sum extends over all the fundamental fields with a non-trivial two-point function with $\Phi_\ih$. The rewriting is useful because now the pole singularity of the original correlators are encapsulated in the Green's function, and the residues are directly related with the amputated amplitude. By comparing with eq.s~(\ref{LSZ0}), (\ref{LSZ1}), taking into account the behavior of the Green's function at the pole (\ref{gf}), we obtain
\bea
&&\langle0|T\left\{\O_1\ldots\O_n\right\}|\Psi_I(\p)\rangle
\mc{N}^{IJ}\mc{M}^\dagger_{j\ih}=\sum_\jh\, \langle \O_1\ldots\O_n[\Phi^\dagger_\jh(p)]_A\rangle \mc{M}_{\jh I} \mc{N}^{IJ}\mc{M}^\dagger_{J\ih}\,,\nn\\
&&\mc{M}_{\ih I}\mc{N}^{IJ}\langle\Psi_I(\p)|T\left\{\O_1\ldots\O_n\right\}|0\rangle
=\sum_\jh\, \mc{M}_{\ih I} \mc{N}^{IJ} \mc{M}^\dagger_{J \jh}
\langle [\Phi_\jh(p)]_A \O_1\ldots\O_n\rangle  \,.
\eea
If the matrix $\mc{M}$ is invertible, and only in this case, the equations above can be turned into closed formulas for the matrix elements, which are thus completely determined by the amputated field correlators. The final result is just the standard one
\bea
&&\langle0|T\left\{\O_1\ldots\O_n\right\}|\Psi_I(\p)\rangle=
\sum_{\ih}\langle \O_1\ldots\O_n\left[\Phi^\dagger_\ih(p)
\right]_A\rangle{\mathcal{M}}_{\ih I}\,,\nn\\
&&\langle\Psi_I(\p)|T\left\{\O_1\ldots\O_n\right\}|0\rangle=
\sum_{\ih}{\mathcal{M}^\dagger}_{I\ih} \langle [\Phi_\ih(p)]_A \O_1\ldots\O_n\rangle\,,
\eea
and it is unaffected by the presence of the non-standard norm matrix $\mc{N}$ in the single-particle scalar products.

\section{Slavnov-Taylor Identities}\label{ST}

Generically we denote as ``Slavnov-Taylor identity'' any relation among the correlation functions in a non-Abelian gauge theory that relies on the BRS symmetry. We will now derive two such identities, for the two-point correlators, which are employed in Section~\ref{mf} of the main text. 

The first step is to define the generating functional $\W$ of connected correlators
\beq
\displaystyle
e^{i\,\W[J,\,K]}=\int{\mathcal{D}}\Phi\,\textrm{exp}\;i\int d^4x\left[
{\mc{L}}+{\mc{L}}_J+{\mc{L}}_K
\right]\,.
\label{genf}
\eeq
In this appendix we will work with the Lagrangian ${\mc{L}}$ of eq.~(\ref{lagtot0}), with the auxiliary field $B$ integrated in to make the BRS symmetry manifest. Therefore the functional integral in the above equation extends all the fields of the theory, including $B$. The proof relies on a judicious choice of the source terms $\mc{L}_J$ and $\mc{L}_K$, which we take to be
\bea
&&{\mc{L}}_I=J_W^{a} \partial^\mu W_{\mu,\,a}+\mt\,\xi\,J_\pi^{a} \pi_{a}+J_{\ob}^{a}\, \ob_{\,a}+J_B^{a} B_{\,a}\,,\nn\\
&&{\mc{L}}_K=K_W^{a} \partial^\mu s(W_{\mu,\,a})+\mt\,\xi\,K_\pi^{a}s( \pi_{a})\,.
\eea
The external sources of the type ``$J$'', which appear in the first term, are coupled to the fundamental fields, the $K$-type ones couple instead to their BRS variations $s(\Phi)$ defined in Table~\ref{table:brs}. Notice that only one scalar source $J_W$ has been introduced for the scalar component $\partial_\mu W^\mu$ of the gauge field, and similarly for its BRS variation. We might have chosen to work with vectorial sources for all the $W^\mu$ components, this would have led to the same final results but with slightly more involved intermediate formulas. Similarly, for simplicity we did not introduce sources for all the other fields and for their BRS variations,  the latter would just complicate the derivation. The choice of the conventional $\mt\,\xi$ factor in front of the Goldstone sources is also dictated by convenience, it is related with the factor that appears in the gauge-fixing functional $f$ defined, as in eq.~(\ref{gff}), by
\beq
\displaystyle
f_a\,=\,\partial_\mu W^{\mu}_a\,+\,\widetilde{m}\,\xi\, \pi_a\,.
\label{gf1}
\eeq

The BRS symmetry leads to a differential equation for the functional $\W$, which is derived as follows. The result of the path-integral is invariant under any redefinition of the field integration variable, we consider in particular an infinitesimal BRS variation, of the form
\beq
\Phi\;\rightarrow\;\Phi+\epsilon\,s(\Phi)\,,
\eeq
where $\epsilon$ is an infinitesimal anti-commuting parameter and the variations $s(\Phi)$ of the fields are reported in Table~\ref{table:brs}. The Lagrangian ${\mc{L}}$ is invariant under the transformation, and also the $K$-source terms in ${\mc{L}}_K$ because the BRS transformations are nilpotent, $s(s(\Phi))=0$. By computing the variation of ${\mc{L}}_J$, keeping in mind that $s(\ob)=B$ and $s(B)=0$, and imposing that the integral stays the same we obtain  
\beq
\displaystyle
\int d^4x\left[
J_W^{a}(x)\dW{K_W^{a}(x)}
+J_\pi^{a}(x)\dW{K_\pi^{a}(x)}
-J_\ob^a(x)\dW{J_B^a(x)}
\right]=0\,.
\label{e30}
\eeq
The one above is the fundamental Slavnov-Taylor relation, from which we will now derive the two identities of Section~\ref{mf}, namely eq.s~(\ref{ST1tx}) and (\ref{last}). Actually, we will not use directly eq.~(\ref{e30}), but its functional derivative with respect to $J_\ob$, which gives
\bea
\displaystyle
\dW{J_B^a(x)}=
\int d^4y\hspace{-15pt}&&\left[
J_W^{b}(y)\ddW{J_\ob^a(x)}{K_W^{b}(y)}
+J_\pi^{b}(y)\ddW{J_\ob^a(x)}{K_\pi^{b}(y)}\right.\nn\\
&&\left.+J_\ob^b(y)\ddW{J_\ob^a(x)}{J_B^b(y)}
\right]\,.
\label{e3}
\eea

The equation above tells us many things about the $B$ field correlators, which are obtained by taking functional derivatives of $\W$ with respect to $J_B$ and eventually setting all the sources to zero. First, it tells us that all the correlators involving only $B$ fields vanish, because $\delta\W/\delta J_B$ is proportional to the other sources. In particular, for the two-point function
\beq
\left.\ddW{J_B^a(x)}{J_B^b(y)}\right|_0=0\,.
\label{eee1}
\eeq
Second, by taking one functional derivative with respect to $J_W$, it allows us to express the mixed $B$-$W$ correlator as
\beq
\left.\ddW{J_W^a(x)}{J_B^b(y)}\right|_0=
\left.\ddW{J_\ob^b(y)}{K_W^a(x)}\right|_0
\,.
\label{eee2}
\eeq
A similar expression could have been derived for the $B$-$\pi$ correlator, but the latter will not be needed for our purposes.

Eq.s~(\ref{eee1}) and (\ref{eee2}) are not yet what we want, for the applications of Section~\ref{mf} we would like to express them in terms of the correlators of the propagating field $W$ and $\pi$, rather than the auxiliary $B$. This is easily achieved, exactly because $B$ is an auxiliary field with trivial equations of motion. Let us go back to the generating functional $\W$ in eq.~(\ref{genf}), and perform an infinitesimal variation on the $B$ variable, $B\rightarrow B+\delta B$, the  result of the integral obviously will not change. The terms of the integrand which depend on $B$ are the $B$ source and the Lagrangian ${\mc{L}}$, in eq.~(\ref{lagtot0}), whose variation is a simple linear polynomial in $B$. By imposing that the integral remains the same we obtain the identity
\beq
\displaystyle
0=\int{\mathcal{D}}\Phi\;
\left\{
\xi B^a(x)+f^a(x)+J_B^a(x)
\right\}
\;\textrm{exp}\;i\int d^4x\,
	\left[{\mc{L}}+{\mc{L}}_J+{\mc{L}}_K\right]
\,,
\eeq
which is immediately rewritten, using eq.~(\ref{gf1}), as
\beq
\displaystyle
\xi\,\dW{J_B^a(x)}+\dW{J_W^{a}(x)}+\dW{J_\pi^{a}(x)}+J^a_B(x)=0\,.
\label{e1}
\eeq

Through the equation above, any correlator involving $B$ can be rewritten in terms of those for the propagating fields $W$ and $\pi$. After some manipulation, making use of eq.~(\ref{e1}) we can turn eq~(\ref{eee1}) into 
\beq
\displaystyle
\left[
\ddW{J_W^a(x)}{J_W^b(y)}+
\ddW{J_W^a(x)}{J_\pi^b(y)}+
\ddW{J_\pi^a(x)}{J_W^b(y)}+
\ddW{J_\pi^a(x)}{J_\pi^b(y)}
\right]_0\hspace{-4pt}=\xi\,\delta^{ab}\delta^4(x-y)\,,
\label{r1}
\eeq
while from eq.~(\ref{eee2}) we obtain
\beq
\displaystyle
\left.\ddW{J_\ob^b(y)}{K_W^a(x)}\right|_0=-\frac1\xi\left[
\ddW{J_W^a(x)}{J_W^b(y)}+\ddW{J_W^a(x)}{J_\pi^b(y)}
\right]_0\,.
\label{r2}
\eeq

We are finally in the position to demonstrate eq.s~(\ref{ST1tx}) and (\ref{last}), we just have to rewrite the functional derivatives in term of field correlators. From eq.~(\ref{r1}) we obtain eq.~(\ref{ST1tx})
\bea
\displaystyle
\partial_\mu^x\partial_\nu^y
\langle W^{\mu,\,a}(x) W^{\nu,\,b}(y)
\rangle
+\mt\,\xi\,\partial_\mu^x
\langle W^{\mu,\,a}(x) \pi^{b}(y)
\rangle&&\nn\\
+\mt\,\xi\,\partial_\nu^y
\langle \pi^{a}(x) W^{\nu,\,b}(y)
\rangle
+\mt^2\,\xi^2\,\langle \pi^{a}(x) \pi^{b}(y)
\rangle&&\hspace{-15pt}=-\,i\,\xi\,\delta^{ab}\delta^4(x-y)\,,
\label{ST1}
\eea
and from eq.~(\ref{r2}), by remembering the explicit form of $s(W)$, we derive eq.~(\ref{last})
\bea
\displaystyle
\xi\langle
\partial_\mu^x\left[
\partial^\mu \o^a
+g\,\varepsilon^{abc}W^\mu_b\o_c
\right]\hspace{-3pt}(x)\ob^b(y)
\rangle=\hspace{-15pt}&&\partial_\mu^x\partial_\nu^y
\langle W^{\mu,\,a}(x) W^{\nu,\,b}(y)
\rangle\nn\\
+\hspace{-15pt}&&\mt\,\xi\,\partial_\mu^x
\langle W^{\mu,\,a}(x) \pi^{b}(y)
\rangle\,.
\label{ST2}
\eea



\begin{thebibliography}{99}
%

\bibitem{Dawson:1984gx} 
  S.~Dawson,
  Nucl.\ Phys.\ B {\bf 249} (1985) 42.
 G.~L.~Kane, W.~W.~Repko and W.~B.~Rolnick,
  Phys.\ Lett.\ B {\bf 148}, 367 (1984);

\bibitem{Kleiss:1986xp}
  R.~Kleiss, W.~J.~Stirling,
  Phys.\ Lett.\  {\bf B182 } (1986)  75.

\bibitem{Kunszt:1987tk}
  Z.~Kunszt and D.~E.~Soper,
  Nucl.\ Phys.\  B {\bf 296} (1988) 253.
  
\bibitem{Borel:2012by}
P.~Borel, R.~Franceschini, R.~Rattazzi and A.~Wulzer,
  JHEP {\bf 1206} (2012) 122
  [arXiv:1202.1904 [hep-ph]].

\bibitem{Chanowitz:1985hj}
  M.~S.~Chanowitz and M.~K.~Gaillard,
  Nucl.\ Phys.\ B {\bf 261} (1985) 379.
  
\bibitem{Horejsi:1995jj}
  J.~Horejsi,
  Czech.\ J.\ Phys.\  {\bf 47} (1997) 951
  [hep-ph/9603321].
    
\bibitem{Bagger:1989fc}
  J.~Bagger and C.~Schmidt,
  Phys.\ Rev.\ D {\bf 41} (1990) 264.
   Y.~-P.~Yao and C.~P.~Yuan,
  Phys.\ Rev.\ D {\bf 38} (1988) 2237.

\bibitem{KO}
  T.~Kugo and I.~Ojima,
  Phys.\ Lett.\ B {\bf 73} (1978) 459; 
  Prog.\ Theor.\ Phys.\  {\bf 60} (1978) 1869; 
  Prog.\ Theor.\ Phys.\  {\bf 61} (1979) 294.
  
\bibitem{Nakanishi:1972sm}
  N.~Nakanishi,
  Phys.\ Rev.\ D {\bf 5} (1972) 1324.
  
\bibitem{Veltman:1989ud}
  H.~G.~J.~Veltman,
  Phys.\ Rev.\ D {\bf 41} (1990) 2294.
  
    \bibitem{Coradeschi:2012iu}
  F.~Coradeschi and P.~Lodone,
  Phys.\ Rev.\ D {\bf 87} (2013) 074026
  [arXiv:1211.1880 [hep-ph]].
  
\bibitem{Kallen:1952zz}
  G.~K\"{a}ll\'en,
  Helv.\ Phys.\ Acta {\bf 25} (1952) 417.
  
\bibitem{Becchi:1974md}
C.~Becchi, A.~Rouet and R.~Stora,
  Commun.\ Math.\ Phys.\  {\bf 42} (1975) 127.
G.~Curci and R.~Ferrari,
  Nuovo Cim.\ A {\bf 35} (1976) 273.
  G.~Bandelloni, A.~Blasi, C.~Becchi and R.~Collina,
  Annales Poincare Phys.\ Theor.\  {\bf 28} (1978) 255.
    P.~A.~Grassi,
  Nucl.\ Phys.\ B {\bf 560} (1999) 499
  [hep-th/9908188].
  

\bibitem{bulkET}
  W.~B.~Kilgore,
  Phys.\ Lett.\ B {\bf 294} (1992) 257.
    H.~-J.~He, Y.~-P.~Kuang and X.~-y.~Li,
  Phys.\ Rev.\ Lett.\  {\bf 69} (1992) 2619;
Phys.\ Rev.\ D {\bf 49} (1994) 4842.
  H.~-J.~He and W.~B.~Kilgore,
  Phys.\ Rev.\ D {\bf 55} (1997) 1515
  [hep-ph/9609326].

\bibitem{SEW}
  V.~S.~Fadin, L.~N.~Lipatov, A.~D.~Martin and M.~Melles,
  Phys.\ Rev.\ D {\bf 61} (2000) 094002
  [hep-ph/9910338].
    P.~Ciafaloni and D.~Comelli,
  Phys.\ Lett.\ B {\bf 446} (1999) 278
  [hep-ph/9809321].
    M.~Ciafaloni, P.~Ciafaloni and D.~Comelli,
  Phys.\ Rev.\ Lett.\  {\bf 84} (2000) 4810
  [hep-ph/0001142].
  
\bibitem{Ferrari:2011bx}
  R.~Ferrari,
  Acta Phys.\ Polon.\ B {\bf 43} (2012) 1735
  [arXiv:1106.5537 [hep-ph]].



\end{thebibliography}
\end{document}